\documentclass[alpha-refs]{wiley-article}
\usepackage{color}
\usepackage{ulem}

\usepackage[T1]{fontenc}                   % Allow for umlaute
\usepackage[utf8]{inputenc}                % Allow for several special characters
\usepackage{caption}
\usepackage[font=normalsize,justification=centering]{subcaption} % for multiple figures combined to one
\usepackage{graphicx}                      % for embedding figures
\usepackage{pdfpages}                      % for embedding pdf figures with \includegraphics
\usepackage{amsmath,bm,amssymb}            % for more mathematical commands and symbols, e.g. align, bm
\usepackage{siunitx}                       % for number and unit formatting
\DeclareSIUnit\dbz{dBZ}
\usepackage{array}                          % for hiding certain columns of a table (column type "H")
\newcolumntype{H}{>{\setbox0=\hbox\bgroup}c<{\egroup}@{}}
\usepackage{dirtytalk}                      % for direct quotation
\usepackage{lineno}

\papertype{Original Article}
\paperfield{Journal Section}

\title{Dynamics and Model Representation of Two Contrasting 
Extreme Precipitation Events in the Sahel}

\author[1]{Souleymane Sanogo}
\author[2]{Marlon Maranan}
\author[2]{Andreas H. Fink}
\author[2]{Beth J. Woodhams}
\author[2]{Peter Knippertz}

\affil[1]{Faculté des Sciences et Techniques (FST),
Université des Sciences des Techniques et de Technologie de Bamako (USTT-B), Bamako, Mali}
\affil[2]{Institute of Meteorology and Climate Research (IMKTRO), Karlsruhe Institute of Technology (KIT), Karlsruhe, Germany}

\corraddress{Marlon Maranan, Institute of Meteorology and Climate Research (IMKTRO), Karlsruhe Institute of Technology (KIT), Karlsruhe, Germany}
\corremail{marlon.maranan@kit.edu}

\fundinginfo{This work was supported by the DAAD Climate Research Alumni and Postdocs in Africa - (climapAfrica) Programme. This research was partly supported by the Deutsche Forschungsgemeinschaft (grant no. SFB/TRR 165, “Waves to Weather”) and conducted within the subproject C2: “Statistical-dynamical forecasts of tropical rainfall”. Prof. Dr. Andreas H. Fink and Dr. Marlon Maranan acknowledge funding from the German Federal Ministry of Education and Research (BMBF) through the WASCAL Research Action Plan 2.0 (WRAP2.0) for FURIFLOOD project (Current and future risks of urban and rural flooding in West Africa – An integrated analysis and eco-system-based solutions; grant No. 01LG2086A) and also from the BMBF project NetCDA (German Academic Network for Capacity Development in Climate Change Adaptations in Africa; grant No. 01LG2301E).}

\runningauthor{Sanogo et al.}

\begin{document}
\maketitle

\clearpage
%\linenumbers
\begin{abstract}
Two extreme flood-inducing precipitation events in two cities in Mali, on 08 August 2012 in San (127 mm) and on 25 August 2019 in Kenieba (126 mm), are investigated with respect to rainfall structures, dynamical forcings, and the ability of the ICOsahedral Nonhydrostatic (ICON) model to represent their evolution. Two sets of experiments with convective parameterization enabled (PARAM) and disabled (EXPLC), both at 6.5 km grid spacing, are conducted for each case. While the (thermo)dynamical fields of the simulations are compared with ERA5 reanalysis data, the rainfall fields are tested against the satellite-based precipitation dataset IMERG by applying the spatial verification methods Fractions Skill Score (FSS) and the Structure-Amplitude-Location (SAL) score. In addition, a spectral filtering of tropical waves is applied to investigate their impact on the extreme events. The most prominent results are: (1) Both cases were caused by organized convective systems associated with a westward propagating cyclonic vortex, but differ in their environmental setting. Although both cases featured an east African wave (AEW), the San case involved convective enhancement along dry Saharan airmasses, whereas the Kenieba case occurred within an unusual widespread wet environment extending deep into the Sahel. (2) Although EXPLC captures the rainfall distribution in the San case better than PARAM, it fails to organize convection in the moisture-laden Kenieba case, which PARAM is capable of simulating. (3) The FSS confirms the case-dependency of the ICON skill. The SAL method hints towards a systematic deficiency of EXPLC to represent the convective organization by producing too many scattered and weak rainfall systems, while PARAM is more effective in converting abundant moisture into excessive rainfall. The results stress the continued need for more research into capturing the complex convective dynamics to better forecast the extremes of Sahelian rainfall.

% Please include a maximum of seven keywords
\keywords{\emph{extreme precipitation}, West Africa, Sahel, ICON model, spatial verification, convection, monsoon}
\end{abstract}
\clearpage
\section{Introduction}  %% \introduction[modified heading if necessary]
\label{sec:introduction}

While the semi-arid zone of West Africa suffered enormously under the great droughts of the 1980s, the rainfall recovery since the 1990s has been associated with frequent episodes of intense rainfall \citep{panthou2014recent,sanogo2015spatio,nicholson2018rainfall}. This trend is expected to continue as the planet warms \citep{westra2014future,kendon2019enhanced,berthou2019larger}. Indeed, the decades that followed the 1990s were marked by the occurrence of extreme precipitation events that have often led to flooding and subsequent impairment of water quality, human health and ecosystems. Those extreme rainfall events affected urban transportation, agriculture, infrastructure and human lives in many countries across the Sahel \citep{tschakert2010floods}, causing considerable socio-economic damages and losses \citep{douglas2008unjust}. According to \citet{ocha2017flood}, more than 11,000 people in Mali alone were affected by floods in the early rainy season of 2017. Pastoral communities were particularly affected with more than 26,000 animals lost across the country. On 16 May 2019, the Malian Red Cross reported a flood event with water levels up to 2.5 meters that affected the capital city Bamako, after a sudden torrential downpour lasting several hours \citep{irfc2017flood}. Such extreme events need to be understood better in order to improve forecast and warning systems, which could crucially prevent damages and minimize risks.

The West African monsoon constitutes a dominating atmospheric feature controlling the large-scale circulation and local precipitation over West Africa. In this regional system, a south–westerly flow is formed between the Gulf of Guinea and the Saharan Heat Low (SHL), which brings moisture to the continent \citep{raji2017radiative, thorncroft2011annual, knippertz2017meteorological}. This mechanism provides the conditions for precipitation in the Sahel from May to October. Many West African rainy seasons in recent decades have been harsh due to the frequent occurrence of flood-inducing rains \citep{njau2011western, paeth2011meteorological, sima2013west, engel2017extreme, fall2020synoptic, nicholson2022meteorological, atiah2023mesoscale}. Some events have attracted the attention of the scientific community, in particular the cases of 01 September 2009 in Ouagadogou, Burkina Faso \citep{engel2017extreme,lafore2017multi,beucher2020simulation}, of 26 August 2012 in Dakar, Senegal \citep{engel2017extreme}, of 12 June 2016 in Abakaliki, Nigeria \citep{maranan2019interactions}, and of 26 August 2017 in Linguère, Senegal \citep{fall2020synoptic}.

The weather systems associated with rainfall events in the Sahel vary throughout the year depending on the stage of the West African monsoon \citep{fink2006rainfall,janiga2014convection,maranan2018rainfall,maranan2020process}. While the peak monsoon period in the Sahel in July and August is dominated by highly organized mesoscale convective systems (MCSs) \citep{fink2003spatiotemporal,zipser2006most,guichard2010intercomparison}, often of squall-line type \citep{diongue2002numerical, redelsperger2002multi, fink2003spatiotemporal}, the transitional months May and October feature rainfall systems that are frequently influenced by tropical-extratropical interactions such as tropical plumes \citep{knippertz2005tropical,knippertz2008dry,rubin2007tropical}. Squall lines, dominating the Sahel during the peak monsoon period \citep{fink2003spatiotemporal}, are fast propagating rainfall features, which can lead to extreme wind gusts but rarely to extreme rainfall. Typically, the associated total rainfall ranges between 20 and 50 mm and hardly exceeds 100 mm \citep{fink2017mean}. However, recent studies have shown that extreme precipitation events over West Africa are often caused by slow-moving MCSs. Such complex systems can be sustained by strong moisture convergence within low-tropospheric vortices, sometimes related to African easterly waves (AEWs), which allow a swift moisture refueling and multiple MCS passages within a short time \citep{engel2017extreme,maranan2019interactions}. For example, for the extreme rainfall event in Abakaliki (Nigeria) in June 2016, such a vortex facilitated the intensification but likely also the deceleration of an MCS, which caused a transformation of the MCS from a fast Sahelian squall line into a slow-moving coastal system \citep{maranan2019interactions}. The subsequent interaction with the vortex led to the maintenance of the MCS through constant and abundant moisture supply along the wet Guinea coast region. In a similar way, the influence of a potential AEW on moisture levels and convergence likely facilitated high precipitation rates during the Linguère case \citep{fall2020synoptic}.

The predictability of extreme rainfall in West Africa using numerical weather prediction (NWP) models has been a long-standing challenge. In general, predictions of daily rainfall occurrence over West Africa with dynamical models have low skill, often not even better than a climatological reference \citep{vogel2020skill}. Investigating the skill of global prediction of rainfall, \citet{vogel2018skill} suspect that the parametrization of convection is a potential cause for the lack of ensemble forecast skill in regions such as the Sahel, which are dominated by MCSs. \citet{pante2019resolving} used the operational weather model of the German Weather Service (Deutscher Wetterdienst, DWD), the ICOsahedral Nonhydrostatic (ICON) model, to compare explicit convection with parameterized convection in simulating diurnal rainfall over West African Sahel. Their results show that explicit convection generally leads to an improved representation of rainfall systems in the Sahel. Whether this translates to a better representation of individual extreme events in the Sahel in ICON is not well known. However, learning about the strengths and weaknesses of a state-of-the-art NWP models in these situations is crucial to identify areas for improvements and the their usefulness for decision-making on an event-to-event basis. This is a key motivation for the work presented in this manuscript.

Here, we investigate two heavy rainfall events with high impacts on the local population that were observed in two Malian cities in August 2012 and 2019, respectively. The two extreme events exhibit distinctly different synoptic-scale dynamics, which requires an in-depth analysis of the development of their synoptic-dynamical features. Specifically, the study seeks (a) to test the capability of explicit and parameterized convection in the ICON model to simulate those rainfall systems and associated dynamic features, and (b) to identify the characteristics of the meteorological environments that favored the development of the extreme precipitation events. The latter is accompanied by investigating the potential impact of equatorial waves \citep{matsuno1966quasi} which are known to modulate and couple with rainfall systems on synoptic timescales \citep[e.g.][]{wheeler1999convectively,roundy2018wave,schlueter2019asystematic,schlueter2019bsystematic}. Furthermore, equatorial waves are related to rainfall extremes in the central Sahel \citep{peyrille2023tropical} and represent a source of enhanced rainfall predictability in the Tropics \citep[e.g.][]{bechtold2008advances,judt2020atmospheric,knippertz2022intricacies}. In the context of West Africa, investigating precipitation extremes is challenging due to the poor availability of appropriate ground-based rainfall data, as station data are not freely accessible in most of the countries in the region and continuous daily and (sub-)hourly rainfall data for periods long enough to identify extreme rainfall events are difficult to acquire. Indeed, to conduct this study, successful efforts were made to obtain daily station rainfall data from the National Meteorological service of Mali.

The paper is structured as follows: Section \ref{sec:data} provides information on the reference data used in this study. Section \ref{sec:methods} describes analysis methods and model experiments. Section \ref{sec:analysis_rainfall_weather} deals with the description and examination of the environmental conditions around the extreme precipitation events on 08 August 2012 in San (referred to as “San case” hereafter) and on 25 August 2019 in Kenieba (“Kenieba case”). Section \ref{sec:evaluation_icon} focuses on the performance evaluation of the ICON simulations compared to observation and reanalysis data with respect to rainfall and dynamical fields. Finally, section \ref{sec:summary} provides a summary and conclusion.

\section{Observational data sets}  %% \introduction[modified heading if necessary]
\label{sec:data}

\subsection{Rain gauges}

As a reference for observed rainfall, daily rain gauge observations are taken from long-term records (i.e., since the early 20th century) of the Karlsruhe African Surface Station Database (KASS-D) \citep{vogel2018skill,schlueter2019asystematic,seregina2019new}. For this study, 50 stations in Mali with at least 90\% of data availability in the period 1960--2019 were considered to identify extreme events, i.e., daily rainfall values larger than the 95th percentile inferred from non-zero daily rainfall amounts over that same period. However, priority is given to extreme events that occurred in the period 2000–-2019 to ensure the availability of IMERG data (see next subsection) for the analysis. Eventually, together with visual inspection of IMERG rainfall maps and quick-looks, the intense rainfall events at the stations San (WMO station ID 61277) and Kenieba (61285) in August 2012 and 2019, respectively, were chosen for this study.

\subsection{IMERG}

In addition to rain gauge records, this study uses rainfall data of the Integrated Multi-satellite Retrievals (IMERG hereafter; version 6B, V6B) for Global Precipitation Measurement (GPM hereafter), the final run of which offered near-global data from June 2000 onwards at the time of compiling this study. A decision was made against the current version 7 (V7) in response to auxiliary tests which showed stronger underestimation tendencies of intense sub-hourly rainfall rates in V7 compared to V6B in the Guinea coast region (not shown). In any case, the relatively fine spatial resolution of 0.1° ×0.1° and high temporal resolution of 30 min \citep{huffman2015nasa, huffman2019v06} makes IMERG highly relevant for investigating the evolution of extreme rainfall events. Moreover, IMERG products have shown good performance in capturing rainfall intensities in arid and semi-arid regions \citep{dezfuli2017validation,gosset2018evaluation,maranan2020process}. The GPM mission, launched on 27 February 2014, comprises an international constellation of satellites to provide the next generation of global observations of precipitation \citep{liu2016comparison}. Rainfall in the IMERG product is estimated by an algorithm that intercalibrates and merges precipitation estimates from the GPM satellite constellation of microwave sensors, microwave-calibrated infrared satellite estimates, and monthly gauge precipitation data.

\subsection{ERA5}

The fifth generation of the European Centre for Medium-Range Weather Forecasts (ECMWF) reanalysis (ERA5), ranging from 1940 until present, is used for analyzing the dynamical fields associated with the selected extreme precipitation events. This global reanalysis is based on the Integrated Forecasting System (IFS) Cy41r2, which became operational in 2016 \citep{hersbach2020era5}. In this study, hourly data with a horizontal grid spacing of 0.25° are used. The variables of interest include relative vorticity, precipitable water, and precipitation rate. We note that ERA5 rainfall is a product of successive short-range IFS forecasts  \citep{lavers2022evaluation} and thus, of the parameterisation schemes in IFS.

\section{Analysis methods and model experiments}  %% \introduction[modified heading if necessary]
\label{sec:methods}

\subsection{ICON model setups} 

\begin{figure}[ht]
	\centering
  \includegraphics[width=12cm]{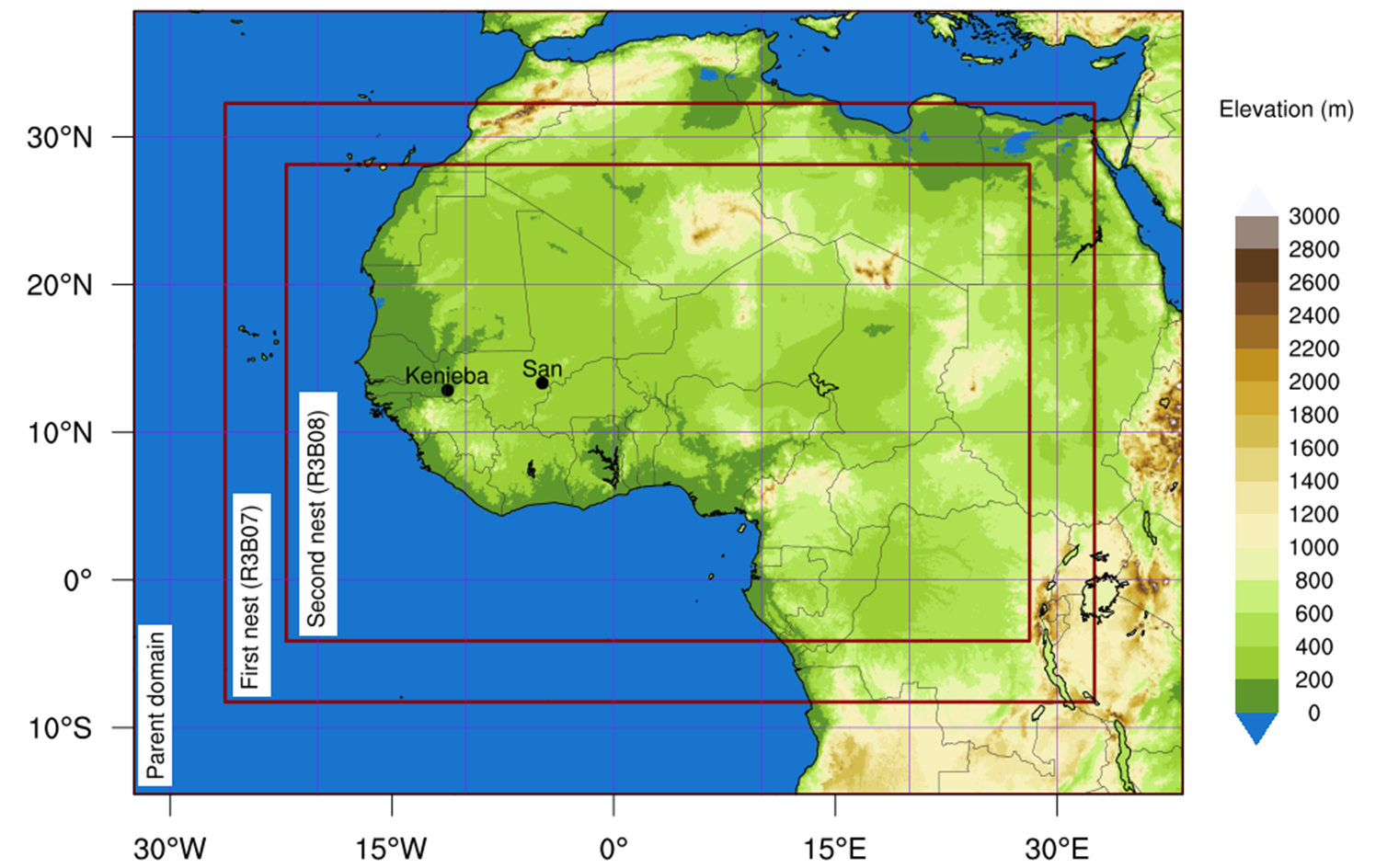}
	\caption{Map showing the terrain elevation (shading) for large parts of Africa, locations of the weather stations of the two Malian cities under study (Kenieba and San, black dots) and the spatial extent of the domains used for the ICON simulations (dark red rectangles). The first nest (R3B07 of the ICON nomenclature \citep{zangl2015icon}) is the coarser domain with a model resolution of $\Delta x=13$ km and a model timestep $\Delta t=120$ s, whereas the second nest (R3B08) is the finer domain with resolutions of $\Delta x=6.5$ km and $\Delta t=60$ s.}
	\label{fig:map}
\end{figure}

The ICON model is a global numerical weather prediction (NWP) model developed jointly by DWD and the Max-Planck Institute for Meteorology in Hamburg (MPI-M) \citep{zangl2015icon}. It has 90 atmospheric levels up to a maximum height of 75 km. The operational ICON setup together with two one-way nests over West Africa is used in this study to assess the capability of both explicit (EXPLC hereafter) and parameterized convection (PARAM hereafter) in representing the two extreme precipitation cases around San and Kenieba (Figure \ref{fig:map}).      

For EXPLC, the parameterisation for deep and shallow convection is switched off and for PARAM, convection is parameterized using the Tiedtke-Bechtold scheme \citep{tiedtke1989comprehensive,bechtold2008advances}. Within the parent domain at 26 km grid spacing, the two nested domains, R3B07 and R3B08, feature successively halved grid spacing of 13 km and 6.5 km, respectively \citep{reinert2018database}. The three domains considered for the present study are shown in Figure \ref{fig:map}. The initial and boundary conditions for the parent domain are taken from ERA5. The lateral boundary conditions for the two inner domains are updated every six hours. Otherwise, the model configuration matches the operational ICON settings with prescribed sea surface temperatures and aerosol properties, as is used in \citet{pante2019resolving}.

\subsection{Spatial verification methods}

The performance of ICON (EXPLC and PARAM) and ERA5 ("test products" hereafter) in representing precipitation is assessed using two spatial verification methods: the Fractions Skill Score (FSS) and the Structure-Amplitude-Location (SAL) score. For both scores, the rainfall fields of all products were aggregated to the same spatial resolution of 0.25°x0.25° to ensure comparability. Neither the FSS nor SAL are computed for the whole simulation domain of ICON but within a 6°x6° box around the individual (subjectively identified) vortex centers in EXPLC, PARAM and ERA5. This is performed in a Lagrangian way such that the boxes in each product shift according to the movement of the vortex at each timestep. In this way, the focus is on the evaluation of the spatial distribution of vortex-related rainfall. Moreover, the effect on the FSS and SAL due to differences in the propagation speed of the rainfall systems is minimized. We note that for IMERG, the centre of the box is always set to that in ERA5. For the sake of simplicity, the term "vortex domain" is used for the above-mentioned box in descriptions of FSS and SAL below.

\subsubsection{Fractions Skill Score (FSS)}

The FSS method, introduced by \citet{roberts2008scale} and further expanded by \citet{roberts2008assessing}, determines the spatial scale at which the test products become skillful regarding precipitation. It relies on the "neighbourhood" approach, i.e., a squared domain around a target grid point with iteratively increasing length of $n$ pixels, over which the rainfall field is verified against a reference (here: IMERG). 

The first step in computing the FSS is the creation of a binary field for each the test products and IMERG where grid points with rainfall values above (below) a threshold are assigned a value of one (zero). In our case, the FSS is used to assess the ability of the models to predict extreme rainfall. Thus, the 95th percentile of all rainfall values (including dry grid points) within the vortex domain was chosen as the threshold, which is determined for each individual product and timestep. A percentile threshold rather than an absolute threshold is applied such that, ideally, both test products and IMERG end up with the same number of grid points with the value of 1 after filtering. To obtain the threshold, the values of all gridpoints were sampled and sorted in ascending order from which the 95th percentile value was determined. Then, all values below this threshold value were set to zero which yields a filtered field of the most intense rainfall for each product and timestep. On average, the 95th percentile corresponds to 53.15 mm in IMERG, 26 mm in EXPLC, 31.85 mm in PARAM, and 25.5 mm in ERA5 at the 0.25°x0.25° resolution.

The next step is to compare the fractions of grid points in IMERG and the test products which exceed the rainfall threshold within a neighbourhood of side length $n$ pixels. The length $n$ is an odd number, with the neighbourhood centered around a grid point. This is an iterative process across all grid points in the vortex domain and with successively larger neighbourhoods up to a size of $n=2N-1$, where $N$ is the number of grid points on the longest side of the vortex domain. The comparison is performed using the mean square error (MSE) between the fractions, where the mean is taken over all grid points. For a neighbourhood with side length of $n$ pixels, the FSS is expressed as (cf., \citealt{roberts2008scale}):

\begin{equation}
    FSS_{(n)} = 1 - \frac{MSE_{(n)}}{MSE_{(n)ref}}
\end{equation}

where a perfect score of $FSS_{(n)}=1$ is only reached when the fraction field of a test product perfectly matches that of IMERG, i.e., $MSE_{(n)}=0$. Here, $MSE_{(n)ref}$ can be considered as the largest possible MSE \citep{roberts2008scale} based on a low-skill reference forecast described in \citet{murphy1989skill}. By ensuring an equal number of threshold-exceeding grid points in IMERG and the test products, the FSS will converge to 1 towards the maximum neighbourhood size. The test products are considered skillful once the FSS reaches a threshold value termed $FSS_{uniform}$. A skillful spatial scale is given by $FSS> FSS_{uniform}=0.5+f/2$, where $f$ is the observed fractional rainfall coverage over the vortex domain. For a deeper and more detailed discussion of the FSS, we refer the reader to \citet{roberts2008scale}. In the present work, the calculation of FSS was performed using the R package “fss”  \citep{pocernich2015package}.

\subsubsection{Structure-Amplitude-Location (SAL) method}

The SAL method \citep{wernli2008sal} consists of a three-component, object-based quality measure comparing rainfall objects between IMERG and the test products where deviations in the structure (S-component), amplitude (A-component) and location (L-component) of precipitation objects are quantified. Again, the 95th percentile value of rainfall at each step is used to delineate the precipitation objects in the vortex domain. In general, all scores contain standardized terms where the S- and A-component, and the L-component range between $[-2, 2]$ and $[0,2]$, respectively. Here, a value of zero indicates a perfect result for all scores. The S-component evaluates the size distribution of precipitation features. Positive values imply too large precipitation objects in the test products. The A-component measures the relative deviation of the domain-averaged precipitation amounts with positive values indicating an overestimation of the rainfall magnitudes in the test products compared to IMERG. Finally, the L-component quantifies two types of location errors: A first term L1 measures the displacement of the centers of mass of the precipitation fields between IMERG and the test products. Through normalization with the largest possible distance of two grid points in the vortex domain, L1 ranges from $[0,1]$. However, since L1 evaluates the precipitation fields as a whole, a second term L2 provides additional information about errors emerging from the individual rainfall objects and quantifies the average distance between the center of mass of the precipitation field and individual precipitation objects. Like L1, L2 is defined within $[0,1]$. Adding L1 and L2 up yields the total L-component. We refer the reader to \citet{wernli2008sal} for a full formulation of all SAL terms.

\subsection{Spectral filtering of tropical waves} \label{sec:sec3_wavefiltering}

To provide insight into the potential role of tropical waves on the two extreme events, the spectral (wavenumber-frequency) wave filtering approach by \citet{wheeler1999convectively} was applied to 6-hourly, normalized rainfall anomalies from IMERG around the periods of interest. The spectral analysis was performed for the latitudinal band 5$^{\circ}$-15$^{\circ}$N and thus centered around the latitudinal position of San and Kenieba. Here, the considered wave types are the Madden-Julian Oscillation (MJO), equatorial Rossby waves (ER), mixed Rossby-gravity waves (MRG), Kelvin waves, and tropical disturbances (including AEWs), the filter settings of which are the same as in \citet{schlueter2019asystematic}. Following the concepts of e.g., \citet{riley2011clouds}, \citet{yasunaga2012differences}, and \citet{schlueter2019bsystematic}, an evaluation of the temporal evolution of the local wave phases and magnitudes during both cases is undertaken by determining the wave-filtered rainfall anomalies and their time derivatives, both of which are standardised by the respective standard deviation value. Within this phase-magnitude spectrum, eight (cyclic) wave phases P1-P8 according to \citet{schlueter2019bsystematic} are defined where P4-P6 indicates the convectively favourable, and P8-P2 the convection-suppressing range. The magnitude of the wave signal is defined by the squared sum of standardised rainfall anomaly and time derivative, visually represented by the distance to the origin. As in \citet{nicholson2022meteorological}, for each of the eight wave phases, the 95\% and 99\% percentile is indicated to assess the extremeness of the wave amplitude. 
\section{Observed rainfall and associated weather systems}  %% \introduction[modified heading if necessary]
\label{sec:analysis_rainfall_weather}

\subsection{Observed station rainfall}

\begin{figure}[ht]
	\centering
  \includegraphics[width=12cm]{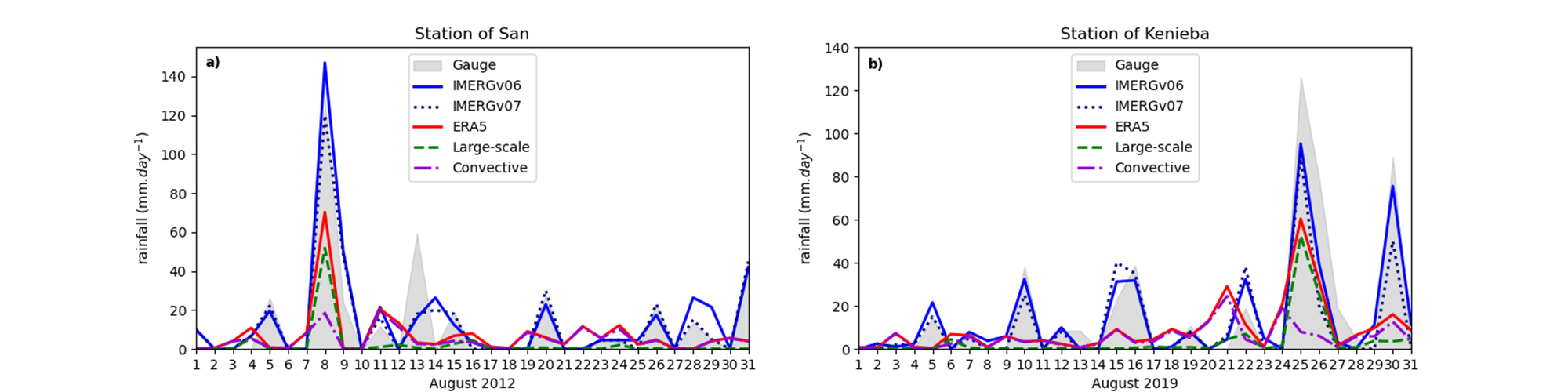}
	\caption{Evolution of daily rainfall (a) at the station of San in August 2012, and (b) at the station of Kenieba in August 2019 from rain gauge data (gray shaded), IMERG estimates (V6B and V7 in solid blue and dashed blue curves, respectively), and total ERA5 rainfall (red curves) and its partitions "grid-scale" (dotted green curves) and "convective" (dashed purple curves) rainfall.}
	\label{fig:timeseries}
\end{figure}

In the following sections, both the San and Kenieba cases are analyzed with respect to the spatio-temporal structures of rainfall in observations. To illustrate the rainfall magnitude of the events, the evolution of daily gauge precipitation at the stations of San in August 2012 (Figure \ref{fig:timeseries}a) and Kenieba in August 2019 (Figure \ref{fig:timeseries}b) is shown (gray shadings). The San case is clearly visible as a distinct spike on 08 August in the rain gauge data (gray shading) with 127 mm, exceeding all other daily rainfall amounts in August 2012 (Figure \ref{fig:timeseries}a). Similarly, the 126 mm peak of the Kenieba case is clear on 25 August 2019 (Figure \ref{fig:timeseries}b). Both cases are extreme in that these rainfall amounts exceed the 99th percentile values (i.e., 80 mm for San and 100 mm for Kenieba) of all recorded wet days within the respective 20-year sample (2000-2019) of the stations. Furthermore, both cases were embedded within marked wet spells, indicated by multiple preceding and succeeding rainfall events. Within seven days, from 08 to 14 August 2012, the station of San recorded 225 mm of rainfall. Also, the period around the Kenieba case was particularly wet with an additional rainfall of 191.8 mm between 26 and 30 August 2019. Thus, approximately 318 mm of rain were accumulated at the station of Kenieba within a span of 7 days.

For comparison, IMERG (V6B and V7 in solid and dashed blue curves, resepctively) and total ERA5 rainfall estimates (red curves) are included in Figure \ref{fig:timeseries}. They are calculated as the 0.5°x0.5° mean values at the four closest grid points to each of the stations of San and Kenieba. The ERA5 estimates are further partitioned into the "grid-scale" (green dotted curves) and "convective" (dashed purple curves) rainfall, which constitute total ERA5 rainfall when combined. For the San case (Figure \ref{fig:timeseries}a), IMERG is able to capture the extreme event, although daily rainfall is overestimated by roughly 20 mm by V6B. A brief comparison between IMERG V6B and V7 shows that V6B tends to peak higher than V7 for extreme daily rainfall. While not necessarily a sign of better quality, V6B overall matches observed intense rainfall closer than V7 according to auxiliary tests, as mentioned in section \ref{sec:data}. In the rest of the manuscript, we will refer to V6B only. Conversely to IMERG, but not surprisingly, ERA5 strongly underestimates the event, which is partly rooted in the poor handling of rainfall by the convection parameterisation scheme in IFS, and ultimately in limitations imposed in by the coarse spatial resolution at 0.25° \citep{rivoire2021comparison,jiang2021evaluation,lavers2022evaluation}. However, relative to their own 20-year samples of daily rainfall, both IMERG and ERA5 values for this event exceed the 99th percentile (not shown). In other words, the San case was extreme within the climatologies of the respective datasets. In addition, the San case is mostly of "large-scale" nature in ERA5, indicating that the area of convective activity was sufficiently large to be resolved by IFS. For the Kenieba case (Figure \ref{fig:timeseries}b), both IMERG and ERA5 slightly underestimate the event on 25 August 2019 but still reach the 99th percentile of August rainfall in their climatologies. Similar to the San case, the majority of the total ERA5 rainfall in the Kenieba case is constituted by the "large-scale" scheme. Being an observational dataset, IMERG overall performed well in estimating magnitude and timing of rainy days.

\subsection{Spatio-temporal evolution of extreme rainfall}

\subsubsection{San case}

\begin{figure}[!ht]
	\centering
  \includegraphics[width=12cm]{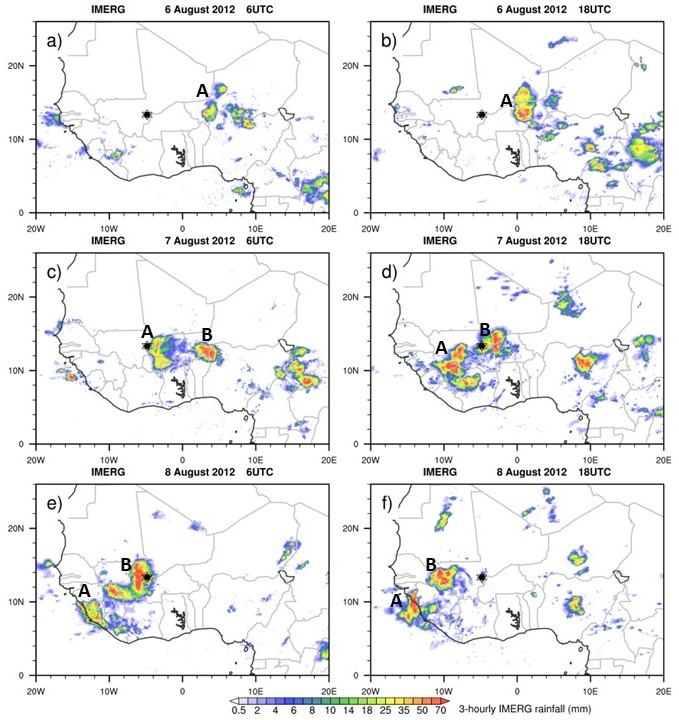}
	\caption{Evolution of 3-hourly IMERG rainfall at important timesteps between 06 and 08 August 2012. The star marker  shows the location of San. The indicated time of day (in UTC) denote the end of the accumulation period, e.g., 03-06 UTC if indicated as 06 UTC. Two major westward propagating rainfall systems are identified and labelled as A and B.}
	\label{fig:imerg_san}
\end{figure}

Figure \ref{fig:imerg_san} highlights the spatio-temporal evolution of 3-hourly IMERG rainfall during the San case at selected timesteps between 06 and 08 August 2012. On 06 August 0600 UTC, i.e. two days prior to the main event, multiple weakly organized rainfall structures are present over southern Niger around 5--10°E (system A in Figure \ref{fig:imerg_san}a), the leading few of which clustered and evolved into a squall line 12 hours later (Figure \ref{fig:imerg_san}b), gradually approaching San. Figure \ref{fig:imerg_san}c depicts two organized rainfall systems. The first reached the station of San around 06 UTC on 07 August 2012. Several convective cores had developed in the wake of this squall line which seemed to reinforce the second rainfall core (system B in Figure \ref{fig:imerg_san}c) arriving at San in the afternoon (Figure \ref{fig:imerg_san}d). This second rainfall core impacted San and surrounding areas for about 12 hours (see system B in Figure \ref{fig:imerg_san}c and d). Ultimately, by early evening of 08 August, this convective cluster has left the San region (Figure \ref{fig:imerg_san}f). 
Overall, the passage of multiple intense convective systems contributed to the extreme nature of the San case. In addition, the event was widespread and reached other synoptic stations in southern Mali causing 56.9 mm in Segou, 61.8 mm in Bamako-Senou, 90 mm in Koutiala and 48.5mm in Sikasso (not shown). With rainfall already occurring the days before (see Figure \ref{fig:timeseries}a), this situation led to severe damage to crop and cattle and destroyed many houses \citep{ocha2012flood}.

\subsubsection{Kenieba case}

\begin{figure}[!ht]
	\centering
  \includegraphics[width=12cm]{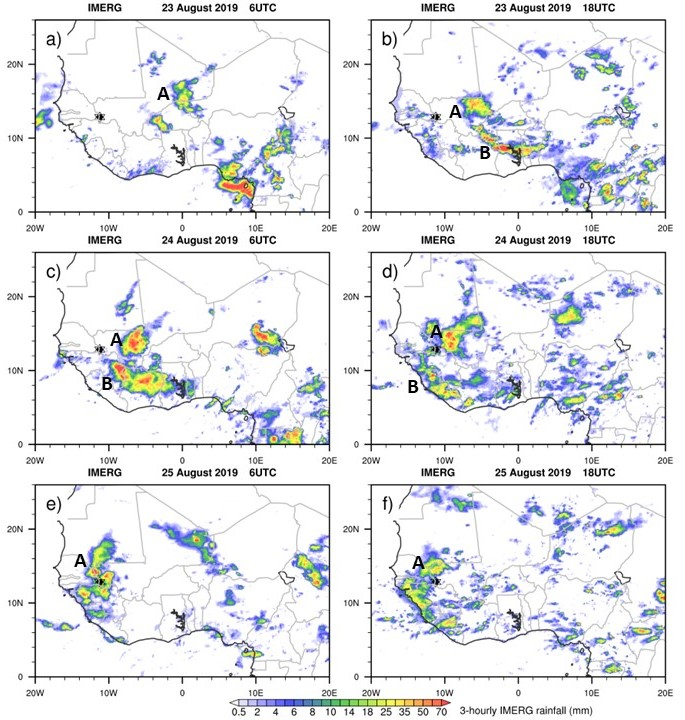}
	\caption{As Figure \ref{fig:imerg_san} but for the Kenieba case, showing selected timesteps between 22 and 25 August 2019.}
	\label{fig:imerg_kenieba}
\end{figure}

Figure \ref{fig:imerg_kenieba} depicts the evolution of 3-hourly IMERG rainfall during the Kenieba case around the 25 August 2019. Two days prior to the event in the morning of 23 August 2019, an organized rainfall system was active around the locality of Hombori (system A in Figure \ref{fig:imerg_kenieba}a). This cluster moved westward and rapidly developed several highly active convective cores (Figure \ref{fig:imerg_kenieba}b), which organized into a curved, zonally oriented band of strong rainfall south of Mali, as well as an intense, circular shaped system just east of Kenieba (systems B and A, respectively, in Figure \ref{fig:imerg_kenieba}c). Overall, the bow-like spatial pattern of strong convection hints at the existence of a rotational flow, i.e., a vortex, in the low- to mid-tropospheric horizontal flow field along which the rainfall systems intensify. Having arrived in the morning of 24 August 2019, the convective cluster subsequently influenced Kenieba for more than 24 hours, as it began to decelerate and eventually stall over the course of 24 and 25 August (Figures \ref{fig:imerg_kenieba}d-f). The meteorological aerodrome (METAR) report at Kenieba airport indicated that the 126 mm on 25 August 2019 were accumulated between 24 August, 12 UTC, and  25 August, 18 UTC. Most of the synoptic stations in the country were affected by this event. 61 mm were recorded at the station of San, 51 mm in Segou, 28 mm in Bamako, 88.5 mm in Koutiala, 79.3 mm in Sikasso, 73.6 mm in Nara, 76.9 mm in Kayes (all not shown). Aside from this, and in contrast to the San case where the convective activity is largely confined to the Sahel in the north, the period around the Kenieba case was also marked by enhanced rainfall activity deep in the Sahara, extending to latitudes beyond 20°N, which, as seen later, was facilitated by a remarkable and widespread northward extension of airmasses with high low-tropospheric moisture.

\subsection{Evolution of associated weather systems}
\label{sec:weather_systems}

This section evaluates the dynamics and synoptic evolution of weather systems at selected timesteps prior to and during the two events.

\subsubsection{San case}  \label{sec:sec6_weather_san}

\begin{figure}[ht]
	\centering
  \includegraphics[width=12cm]{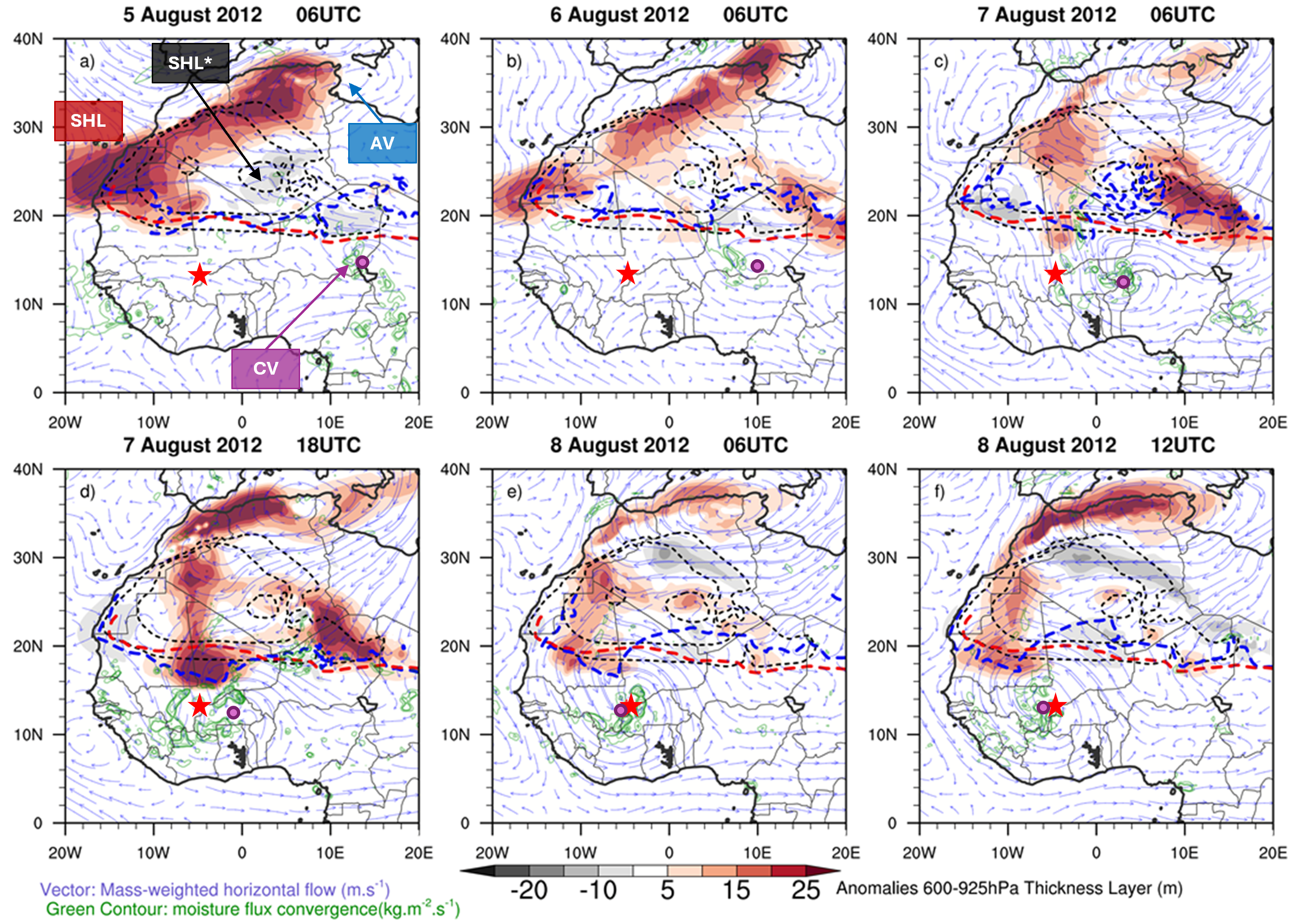}
	\caption{Evolution of weather systems at selected timesteps prior and during the San case from 05 to 08 August 2012 over West Africa. The red star and the purple circled dot show the positions of San and the center of the rain-bearing cyclonic vortex, respectively. Vectors denote 925–-600 hPa mass-weighted flow, shadings indicate anomalies of the 925–-600 hPa layer thickness (used as proxy for the location of the heat low) with respect to the 1981–-2010 long-term monthly mean (black dash contour), blue and red lines indicate the actual and climatological ITD position, respectively, defined by the 14°C isodrosotherm at 2m, and moisture flux convergence is represented by green contours at $-12\times10^{-3}$ to $-0.5\times10^{-3}$ kg m$^{-2}$ s$^{-1}$. In a), SHL, SHL*, AV, and CV denote the Saharan heat low, the climatological SHL in the month of August, the anticyclonic vortex over the Mediterranean and northern Africa, and the cyclonic vortex, respectively.}
	\label{fig:weather_san}
\end{figure}

Figure \ref{fig:weather_san} depicts the spatio-temporal evolution of the strongest moisture convergence (green contours), the mass-weighted mean of horizontal wind between 925 and 600hPa (arrows) and the relative location of the heat low defined as layer thickness as well as the intertropical discontinuity (ITD) prior to and during the San case. The SHL location is determined using the low-level thickness layer between 925 and 700hPa as in \citet{lavaysse2009seasonal}. The anomalies (colored shadings) from the long-term August mean of the low-level layer thickness (i.e., climatological SHL position; black dashed contours and center labelled as SHL*) depict the change in the location of the SHL, where anomalously high positive values (i.e., red shading) indicate its position.

A few days prior to the San case, the SHL is located anomalously to the west at the Atlantic coast (Figure \ref{fig:weather_san}a,b). At its climatological position over south-eastern Algeria (5°E, 25°N; labeled SHL*), the layer thickness is lower than normal, which appears to be due to ventilation from an anticyclone over the Mediterranean Sea (labeled AV). At the same time, the mass-weighted flow indicates the existence of a weak cyclonic circulation over Lake Chad (15°E, 15°N; labeled CV), from where it starts moving westward. By 07 August (Figure \ref{fig:weather_san}c), one day prior to the San case, this vortex has visibly intensified over western Niger (4°E, 12°N), where it exhibits pronounced moisture convergence in its center. Furthermore, together with a strengthened anticyclone over the Mediterranean Sea, a continuous north-easterly flow extending from the Mediterranean Sea to the western flank of the cyclonic vortex is established. Over the course of 07 August, this leads to a southward bulge of the ITD to the northwest of the cyclonic vortex, which is likely due to a southward advection of dry airmasses related to the SHL. The consequence is a local maximum of layer thickness anomaly just north of San (5°W, 18°N) (Figure \ref{fig:weather_san}d). This is the time when the station (red star) was passed by multiple intensified convective cores in the wake of the squall line (cf., Figure \ref{fig:imerg_san}d). The southward intrusion of dry air and the subsequent lifting of moist airmasses probably facilitated the intensification of these convective cores, similar to the effect of drylines known in the midlatitudes \citep[e.g.][]{johnson2018effects}. In this regard, \citet{klein2020dry} outlined a mechanism where the convergence of moist monsoon airmasses and a southward excursion of drylines and the ITD, the latter leading to dry anomalies in soil moisture and stronger heat fluxes, creates enhanced convective instability which facilitates the intensification or even the generation of MCSs. Indeed, pronounced soil moisture gradients prevail during the San case along the storm path (not shown) which overall might have promoted the initiation and maintenance of these successive convective cores. In any case, the cyclonic vortex further intensifies and likely causes the convective systems to strengthen as well (cf., Figure \ref{fig:imerg_san}e,f) while further providing moisture as seen by pronounced moisture convergence in the vortex center during the San case (Figure \ref{fig:weather_san}e,f). 

\subsubsection{Kenieba case} \label{sec:sec6_weather_kenieba}

\begin{figure}[!htbp]
	\centering
  \includegraphics[width=12cm]{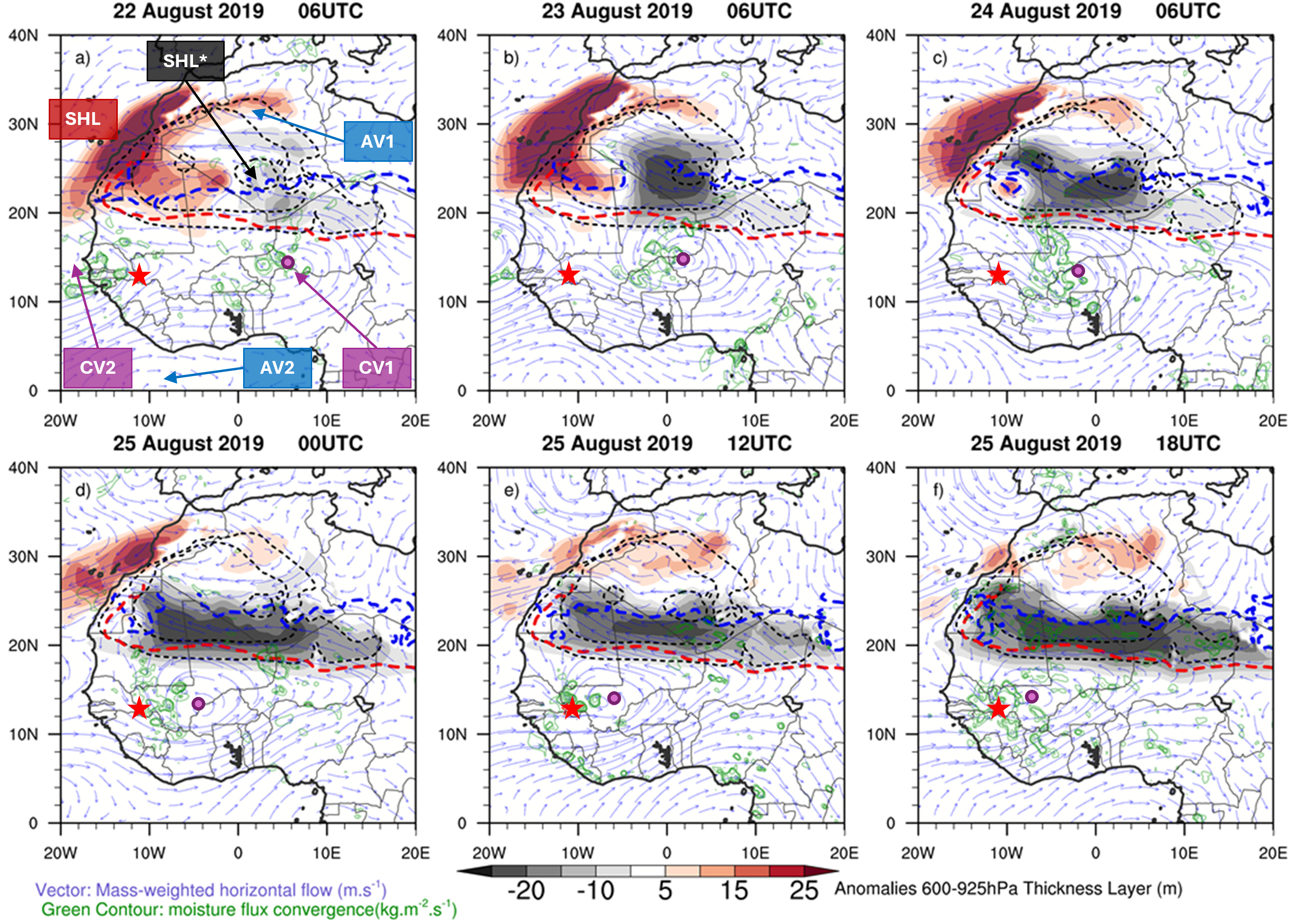}
	\caption{Similar to Figure \ref{fig:weather_san}, but around the Kenieba case from 22 to 25 August 2019. In a), AV1 and CV1 refer to the relevant anticyclonic and cyclonic vortices, respectively, over the continent, while AV2 and CV2 refer to the vortex couple over the eastern Atlantic.}
	\label{fig:weather_kenieba}
\end{figure}

Similar to the San case, the Kenieba case occurred during a period with a weakening of the SHL, but to a considerably stronger extent. Figure \ref{fig:weather_kenieba} shows the environmental conditions up to three days prior to the event on 25 August 2019. On 22 August, the SHL is located at the coast of northwestern Africa after being pushed westward by a weak negative anomaly in the 600-925hPa thickness layer over southern Algeria (Figure \ref{fig:weather_kenieba}a). The latter is sustained by a cyclonic vortex (labeled CV1) over southern Niger (red circle, 6°E, 13°N) and a weak anticyclone (labeled AV1) over northern Algeria (7°E, 33°N), both advecting cooler and moister air into the Sahara. At this point, a pronounced northward shift of the ITD to 25°N (blue dashed line), far from its climatological position at 20°N (red dashed line), is evident in almost the entire domain, which emphasizes a deep moistening of the southern Sahara, stronger than during the San case. This is likely facilitated by a north-south-aligned vortex couple over the Atlantic (labeled AV2 and CV2), which establishes a low-level, westerly “conveyor belt” of moist and cool air over the Guinea Coast region. Eventually, these airmasses are picked up by the cyclonic vortex over Niger and transported northward into the Sahara on its eastern flank. Over the days preceding the Kenieba case, the negative layer thickness anomaly intensifies and extends westward due to the strengthening of the cyclonic and anticyclonic vortices over land (Figure \ref{fig:weather_kenieba}b) and their gradual westward displacement (Figure \ref{fig:weather_kenieba}c-f), which leads to an erosion of the SHL. By the time of the event (Figure \ref{fig:weather_kenieba}d-f), the entire Guinea Coast region is under the influence of a southwesterly flow, driven by the cyclonic vortex now located east of Kenieba at 5°W, 13°N. Thus, Kenieba (red star) is located at its western flank and in an area of enhanced moisture flux convergence (green contours), which likely played a key role in the development of extreme rainfall over the station. Furthermore, it is suggested that the pronounced moistening of the southern Sahara prevented strong evaporation to an extent that allowed the vortex to wrap sufficient moisture to its downstream flank for the build-up of enhanced precipitation.

\subsection{Role of equatorial wave activity}

\begin{figure}[!htbp]
	\centering
  \includegraphics[width=12cm]{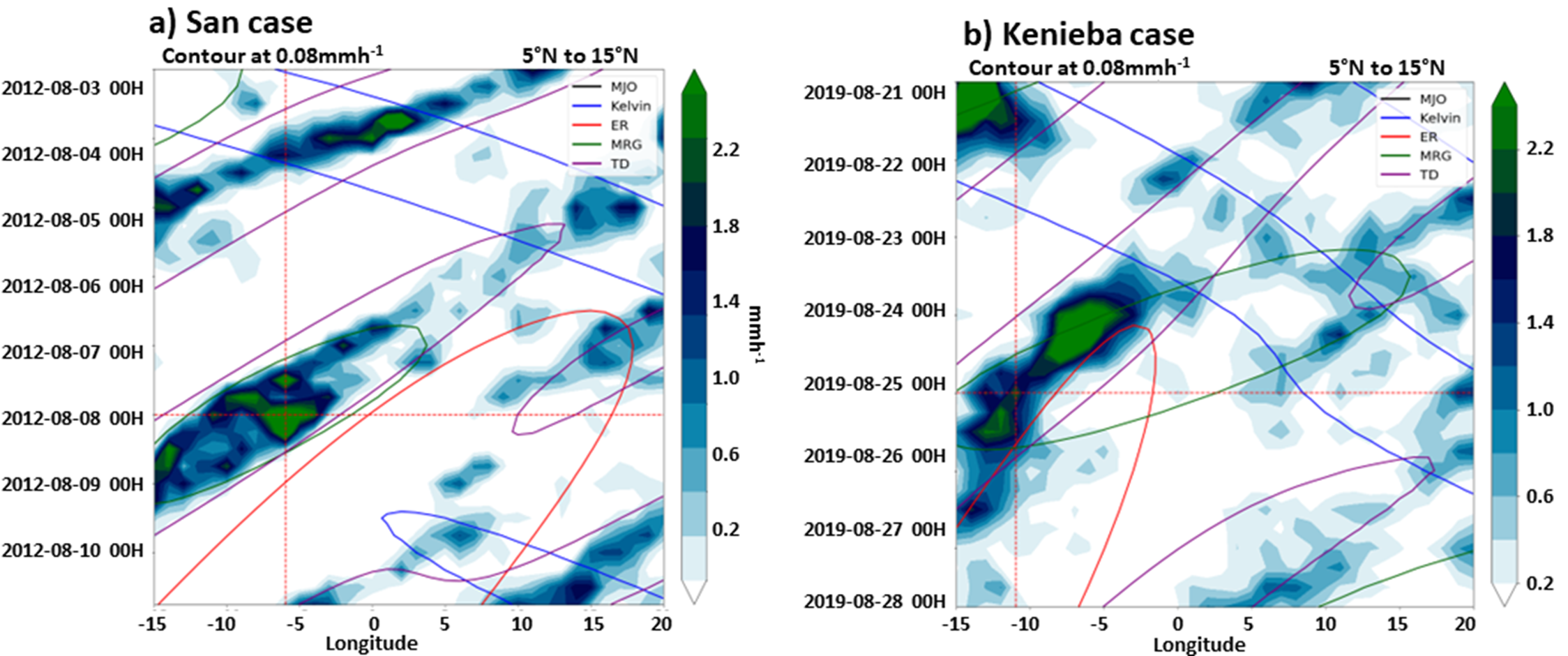}
	\caption{Hovmöller diagram of wave-filtered six-hourly IMERG rainfall within the 5°–15°N latitude band for the periods (a) 03 to 10 August 2012 around the San case, and (b) 21 to 27 August 2019 around the Kenieba case. Vertical and horizontal red lines indicate the position of the stations and the event time, respectively. Contours denote a wave-filtered rainfall rate above 0.08 mm h$^{-1}$ with colours indicating different wave types. The coloured shadings indicate the meridionally averaged six-hourly IMERG rainfall. The diagram was created using the wave-number frequency filtering concept of \citet{wheeler1999convectively} and parameters in \citet{schlueter2019asystematic}.}
	\label{fig:waves_hovmoller}
\end{figure}

Regarding the understanding of the dynamical evolution of extreme Sahelian rainfall, highly organized convective systems such as those shown in this manuscript, as well as in other studies (e.g., \citet{engel2017extreme}), are often associated with the presence of AEWs, but have also been identified to link with other equatorial waves (e.g., \citet{lafore2017multi}). Following the application of the spectral wavenumber-frequency filtering approach described in section \ref{sec:sec3_wavefiltering}, an overview of equatorial wave activity around both the San and Kenieba cases is given in Fig. \ref{fig:waves_hovmoller}, where the Hovmöller diagram displays the propagation of convectively favourable (i.e., positive) wave phases (coloured envelopes) during the periods of interest. The respective time and date of the events at the location of the San and Kenieba stations are indicated by the intersection of the horizontal and vertical red lines in this longitude-time space.

The previously analysed low-tropospheric vortices are clearly identified as westward propagating tropical disturbances in both cases (TDs, purple contours in Figs. \ref{fig:waves_hovmoller}a,b) which move alongside the rainfall signals causing the extreme events (shaded features). Each of the TD passages during the events were preceded by another positive TD phases by roughly four days. This falls well within the periodicity of AEWs \citep[e.g.,][]{burpee1972origin,fink2003spatiotemporal} which supports presumptions that both extreme rainfall cases, and the emergence of low-level vortices therein, were driven by AEWs. Furthermore, a common feature in both cases is an embedded, westward moving positive MRG phase (green contours), which appears to coincide with an intensification of convective activity. The concept of AEW-MRG interactions and mutual amplification abilities over Africa is not unknown and has been the subject of extended evaluation in \citet{yang2018linking} and \citet{cheng2019two}. However, it shall be noted that since the ranges of frequency and wave number for the filtering of TD and MRG partially overlap (see \citet{schlueter2019asystematic}), signals of both waves are not entirely distinguishable. Therefore, it is possible that TDs project onto MRGs and vice versa, which might be the case for the San event (Fig. \ref{fig:waves_hovmoller}a). Regarding other waves types, the development of the TD and MRG in the Kenieba case (Fig. \ref{fig:waves_hovmoller}b) could have benefitted from the existence of an eastward propagating Kelvin wave (blue contour), which intersects the waves already further upstream around and east of the zero meridian, respectively. Although also existent in the San case (Figs. \ref{fig:waves_hovmoller}a), a potential influence of the Kelvin wave on the two waves prior to the event is less obvious, but might have facilitated the intensification of the TD. Overall, it appears that both cases were largely influenced by the fast equatorial wave modes. Out of the slower waves, only ER signals (red contours) are identified, which might have had an influence at least during the Kenieba case (Fig. \ref{fig:waves_hovmoller}b). Overall, the analyses is consistent with the statistical results discussed in \citet{peyrille2023tropical} that document the prevalent role of AEWs for extreme precipitation events in the region, with the Kelvin wave enhancing the convection and the ER being a potential large-scale driver.

\begin{figure*}[!htbp]
        \centering
        \begin{subfigure}[b]{0.475\textwidth}
            \centering
            \includegraphics[width=\textwidth]{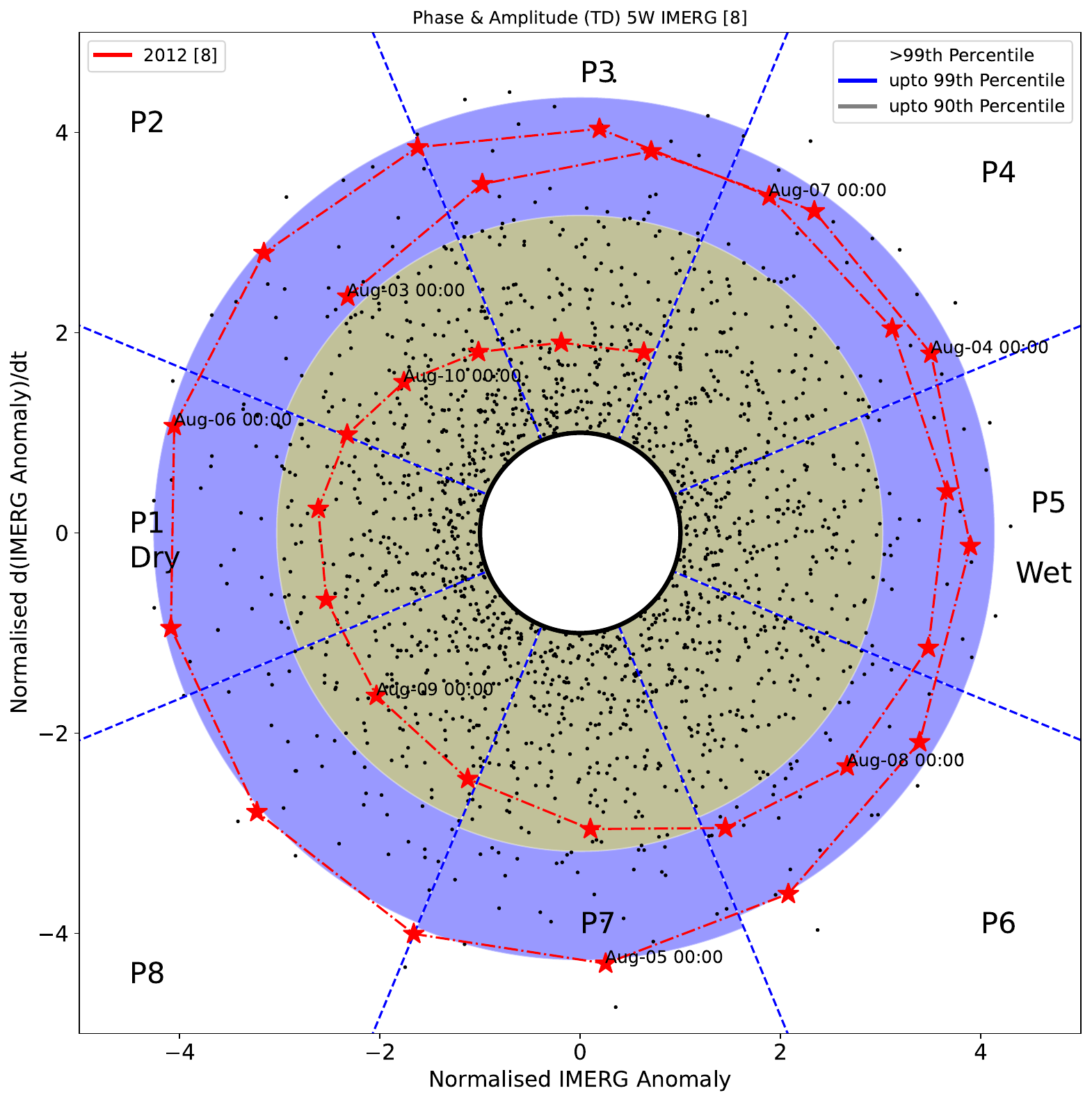}
            \caption[San_TD]%
            {{\small San TD}}    
            \label{fig:amp_phase_san_td}
        \end{subfigure}
        \hfill
        \begin{subfigure}[b]{0.475\textwidth}  
            \centering 
            \includegraphics[width=\textwidth]{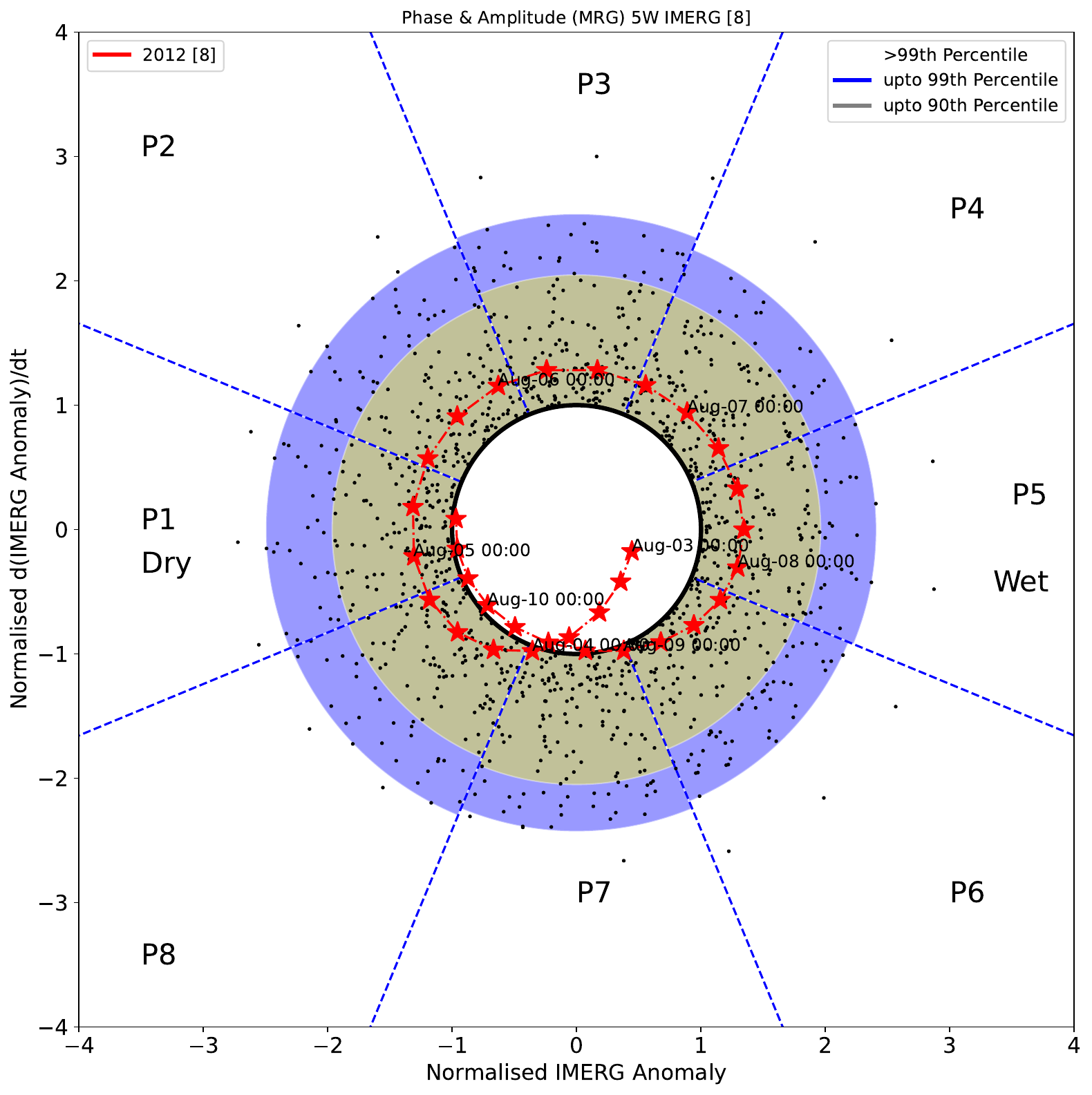}
            \caption[San_MRG]%
            {{\small San MRG}}    
            \label{fig:amp_phase_san_td}
        \end{subfigure}
        \vskip\baselineskip
        \begin{subfigure}[b]{0.475\textwidth}   
            \centering 
            \includegraphics[width=\textwidth]{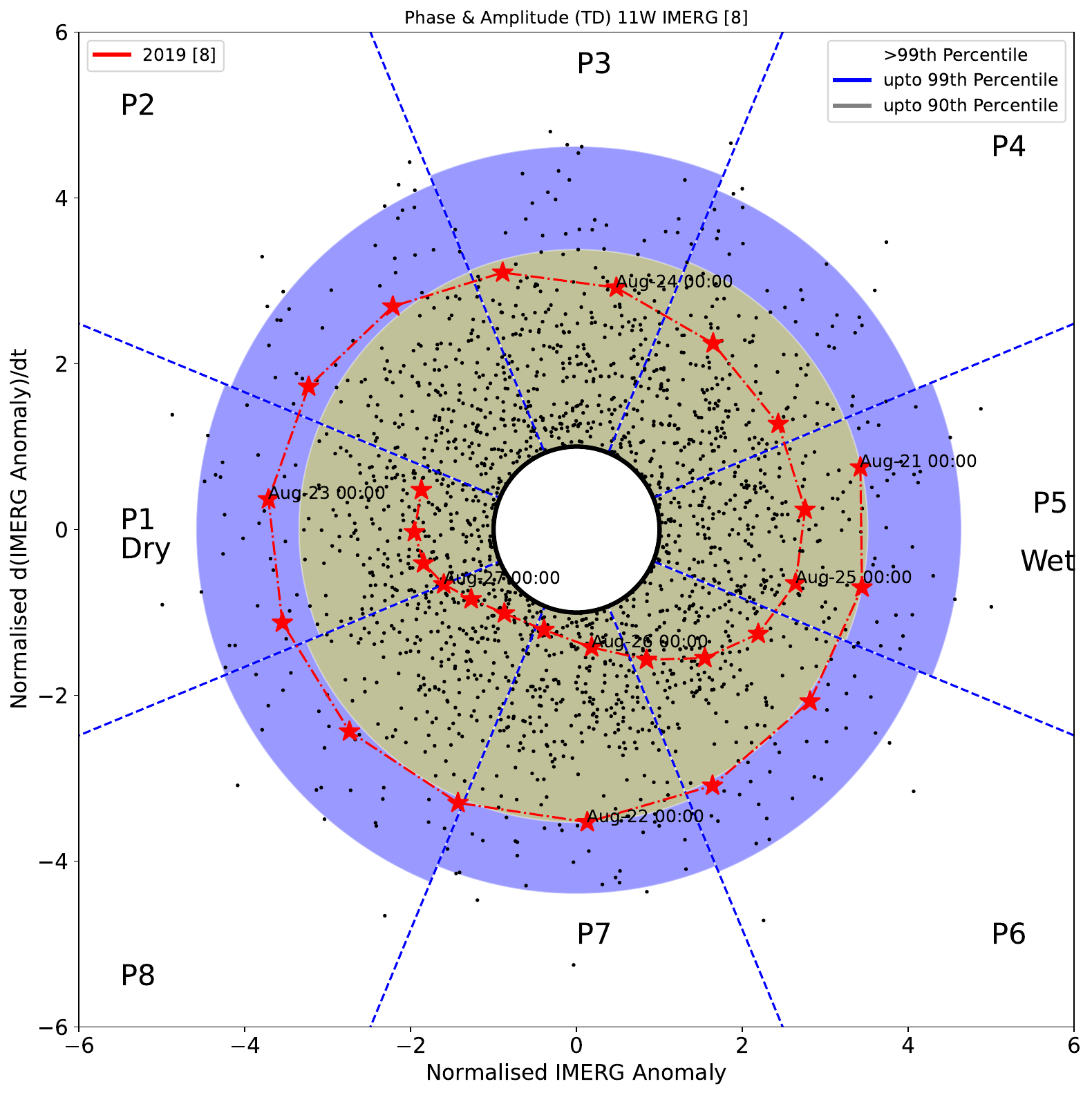}
            \caption[Kenieba_TD]%
            {{\small Kenieba TD}}    
            \label{fig:amp_phase_kenieba_td}
        \end{subfigure}
        \hfill
        \begin{subfigure}[b]{0.475\textwidth}   
            \centering 
            \includegraphics[width=\textwidth]{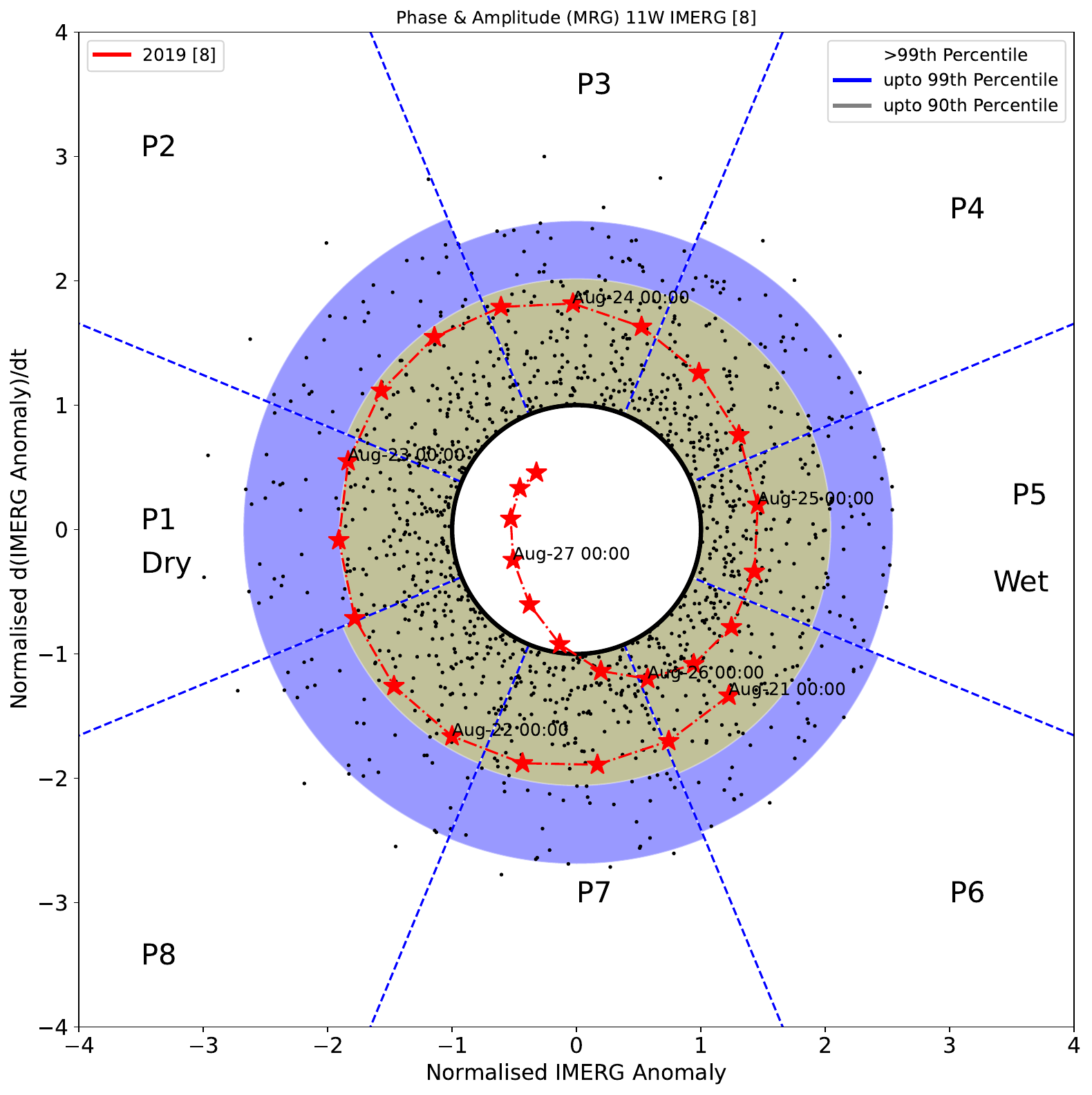}
            \caption[Kenieba_MRG]%
            {{\small Kenieba MRG}}    
            \label{fig:amp_phase_kenieba_mrg}
        \end{subfigure}
        \caption[ The average and standard deviation of critical parameters ]
        {\small Phase diagram of the local (i.e., based on the station longitude) wave activity, defined by the standardised IMERG rainfall anomaly (x-axis) and its time derivative (y-axis). The spectrum is subdivided into eight phases, where P1 and P5 denote the peak dry and wet phase, respectively. The red star markers indicate each timestep during the respective cases, which form the red trajectory in the phase spectrum. Timesteps with a standardised value $<$1, i.e., being within the white center circle, denote an inactive wave. For each of the eight wave phases, the golden and blue shaded areas encompass the sample (black dots) of local wave amplitudes up to the 90th and 99th percentile, respectively. Black dots outside the boundary of the blue area indicate the strongest 1\% amplitudes. Note the different axis range, i.e. range of wave magnitudes, between the TD and MRG plots.} 
        \label{fig:amp_phase_total}
\end{figure*}

Evaluating the most relevant waves in both cases in more detail, Fig. \ref{fig:amp_phase_total} shows the prevailing wave phase and standardised wave amplitude of TD and MRG during the San (Figs. \ref{fig:amp_phase_total}a,b) and Kenieba case (Figs. \ref{fig:amp_phase_total}c,d) from an Eulerian perspective. The local wave phase P1-P8 at the longitude of the San and Kenieba stations, respectively, can be inferred by the red trajectory within the phase room where the local wave amplitude is further indicated by the radial distance from the plot center. Moreover, the shaded areas in each wave phase encompass the sample of local wave amplitudes for the period 2001-2019 (August data only) where the upper boundary of gold and blue area delineate the 90th and 99th percentile, respectively. If the trajectory is located inside of the center circle, i.e., having a standardised amplitude of less than 1, the wave is considered to be inactive, otherwise active.

First of all, based on the extent of the shaded areas, it is evident that can TDs (Figs. \ref{fig:amp_phase_total}a,c) reach higher standardised amplitudes than MRGs (Figs. \ref{fig:amp_phase_total}b,d), which reflects how frequently intense rainfall in Sahelian MCSs is coupled with the existence of TDs. 
Over the course of the analysis period of the San case, the San region is under the influence of two full TD/AEW cycles, the second of which belongs to the actual San event (7-8 August, Fig. \ref{fig:amp_phase_total}a). While being in the convectively favourable phases P4-P6, it is evident that the amplitudes of the TD signal in both cycles are among the strongest 10\% of the entire sample, even approaching the 99th percentile. While being considered as active during the event, the magnitude of the MRG signal does not particularly stand out (Fig. \ref{fig:amp_phase_total}b) but likely was a facilitator for the intensification of the MCS. However, as insinuated earlier, it is debatable whether this MRG signal is indeed a stand-alone wave and not just an additional projection of the TD. Nonetheless, the enhanced wave activity supports the presumption of an increased role of tropical waves in the San case. Likewise, the Kenieba region is under the influence of an active TD and MRG in P5 during the event (25 August, Figs. \ref{fig:amp_phase_total}c,d). Covering the same amount of days as the San case in the phase diagrams, the most apparent difference to the San case is the weaker but also slower TD signal (i.e., 1.5 TD cycles). All the same, it has led to a similar daily rainfall amount as in the San case (126 mm and 127 mm in the Kenieba and San case, respectively). Therefore, while amplitude, propagation speed, and the superposition of wave signals generally drive the regional-scale character of convective activity, the resulting rainfall amounts on the local scale are still subject to the meso-scale evolution of individual cloud systems.

\section{Evaluation of ICON simulations}
\label{sec:evaluation_icon}

\subsection{Comparison of rainfall fields between observations and ICON simulations}

\begin{figure}[!htbp]
	\centering
  \includegraphics[width=12cm]{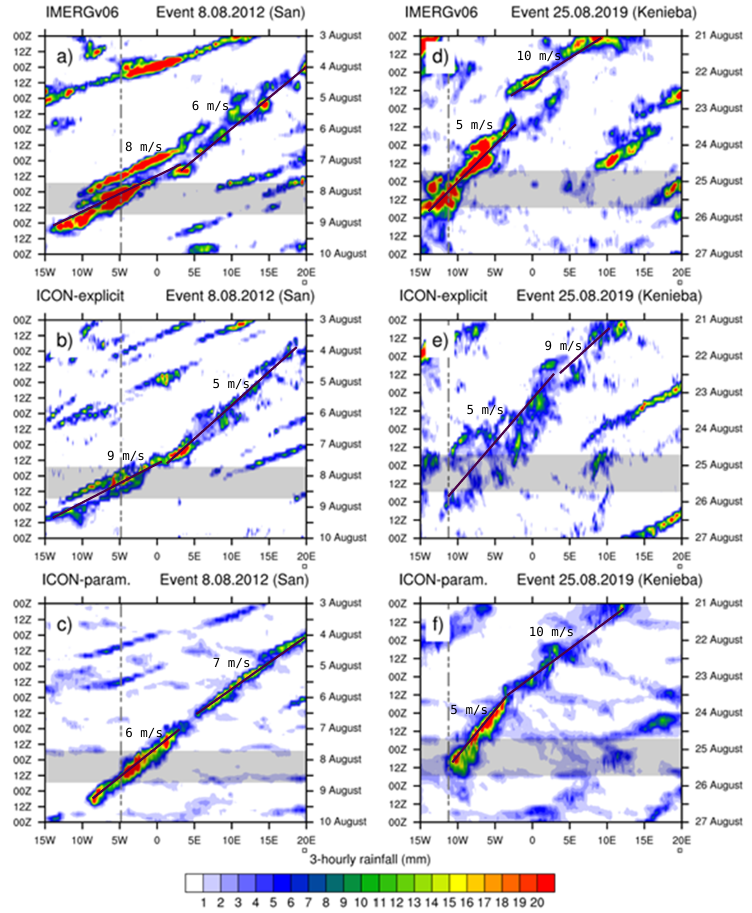}
	\caption{Hovmöller diagrams showing 3-hourly accumulated rainfall from IMERG (a,d), EXPLC (b,e) and PARAM (c,f). The panels on the left side depict the extended period prior and during the San case (08 August 2012) and panels on the right side prior and during the Kenieba case (25 August 2019). Rainfall is averaged over a 5° latitude band centered over the coordinates of San and Kenieba, respectively. The vertical lines indicate the longitudinal positions of San and Kenieba. The horizontal, gray-shaded band refers to the time when the rains were active at San and Kenieba, respectively, according to IMERG. The black lines denote estimated velocity slopes with their respective values.}
	\label{fig:rr_hovmoller}
\end{figure}

Both the San and Kenieba cases were simulated with ICON, which was initialized several days prior to each verification date in order to accommodate important dynamical features discussed in the previous section. As detailed in section \ref{sec:methods}, two experiments for each case are conducted: ICON with convective parameterization enabled (PARAM) and disabled (EXPLC). To facilitate the comparison, 3-hourly rainfall in IMERG and the ICON experiments are displayed in Hovmöller diagrams for the San (Figure \ref{fig:rr_hovmoller}), left column) and the Kenieba case (Figure \ref{fig:rr_hovmoller}, right column). Here, rainfall was meridionally averaged within a 5° band with the center latitude at the coordinates of San and Kenieba, respectively.

\subsubsection{San case} \label{sec:sec4_san}

During the period of the San case (gray shading), the longitude of the station (vertical dashed line) experiences the passage of two major rainfall streaks in IMERG (Figure \ref{fig:rr_hovmoller}a), which reflects both the squall line and the intense convective cells in its wake (see Figure \ref{fig:imerg_san}). EXPLC is able to capture the entire westward propagation of the convective cells, although at considerably lower intensity and a delay in arrival of roughly 12 hours with respect to the first streak (Figure \ref{fig:rr_hovmoller}b). Furthermore, a more detailed inspection of the precipitation fields reveals that the spatial extent of the rainfall systems in EXPLC are underestimated compared to IMERG (not shown), which partly explains the weaker signals. Nonetheless, signs of consecutive passages of convective systems are represented in EXPLC. Their propagation speed is also comparable to the estimation in IMERG (9 ms$^{-1}$). The convective activity prior to the San case is generally overestimated but somewhat misses the intensity of the rainfall event around 04 August 2012. The simulation with PARAM likewise captures the westward propagation of rainfall systems (Figure \ref{fig:rr_hovmoller}c), which suggests that the evolution of the San case was steered by large-scale dynamics. The San region experiences the passage of a single but more intense rainfall streak in PARAM with lower translation velocity (7 ms$^{-1}$) compared to EXPLC (8 ms$^{-1}$), the structure of which shows less “noise”.

\subsubsection{Kenieba case} \label{sec:sec4_ken}

The Kenieba case is marked by westward propagating systems that become organized and intense on 24 August 2019 (Figure \ref{fig:rr_hovmoller}d). Eventually, this manifests in intense signals at the longitude of Kenieba, lasting for entire 25 August 2019. Although convective activity associated with the Kenieba case along with its westward propagation is visible in EXPLC, the rainfall signals never reached the magnitudes of IMERG (Figure \ref{fig:rr_hovmoller}e). This is because the rainfall systems in EXPLC failed to organize into convective clusters (not shown), in stark contrast to developments seen in IMERG in Figure \ref{fig:imerg_san}c. The rainfall signals in the Hovmöller diagram appear very scattered and seem to influence the Kenieba region beyond the period of the Kenieba case. PARAM on the other hand shows strong similarities to the convective development in IMERG (Figure \ref{fig:rr_hovmoller}f). The rainfall intensification becomes apparent around the same time. However, the subsequent westward propagation in PARAM occurs at a slightly lower pace (at a velocity of 9 ms$^{-1}$). Moreover, PARAM seems to lack the capability to sustain convection beyond 10°W which, interestingly, is also apparent in the San case. While a deeper analysis is beyond the scope of this study, it can be speculated whether the influence of the topography of the Guinea Highlands plays a role in the loss of convective organization in PARAM.

Overall, the San and Kenieba cases show rather contrasting outcomes with respect to the performance of explicit convection. Particularly the Kenieba case demonstrates that EXPLC does not necessarily lead to superior results compared to PARAM with respect to the organization of large convective activity. However, EXPLC generally appears more capable of reproducing smaller-scale convective systems and simulating their westward propagation.

\subsection{Spatial verification of ICON precipitation fields}  %% \introduction[modified heading if necessary]
\label{sec:spatial_verification}

This section focuses on the comparison of spatial features of the three-hourly rainfall fields in the ICON simulations to IMERG estimates during both the San and Kenieba events using FSS and SAL (see section \ref{sec:methods}). In the following, they are applied to the timesteps shown in Figures \ref{fig:imerg_san} and \ref{fig:imerg_kenieba}, i.e., on 3-hourly rainfall. The rainfall fields of ERA5 are additionally evaluated in this analysis to provide the perspective of another product with parameterized convection.

\subsection{San case}

\begin{figure}[h]
	\centering
  \includegraphics[width=12cm]{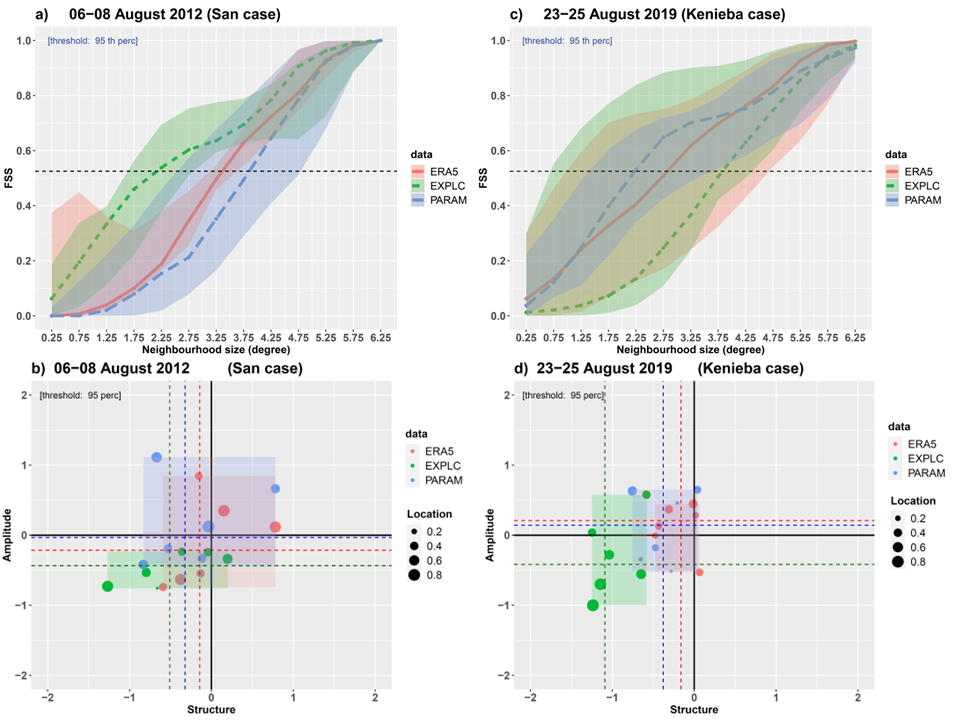}
	\caption{Spatial verification methods FSS (top row) and SAL (bottom row) applied on the San case (left column, a+b) and the Kenieba case (right column, c+d)  with IMERG as the reference. Both FSS and SAL are calculated from the ensemble of timesteps indicated in Figures \ref{fig:imerg_san} and \ref{fig:imerg_kenieba} for the San and Kenieba case, respectively. The ensemble of FSS curves (solid lines: median; shaded area: extent of the minimum-maximum) is plotted as a function of neighbourhood size (in degrees) for EXPLC (green), PARAM (blue) and ERA5 (red). The horizontal dashed line denotes $FSS_{uniform}$. For SAL, the S- and A-components (Structure and Amplitude) are represented on the abscissa and ordinate, respectively, the L-component (Location) is quantified through the size of the points. The colour-shaded rectangles denote the extent of the minimum-maximum in the S-A-component space while the median is found at the intersection of the dashed lines. The SAL median values for both cases are also indicated in Table \ref{tab:fss_sal}}
	\label{fig:fss_sal}
\end{figure}

For the San case, Figure \ref{fig:fss_sal}a depicts the FSS for EXPLC (green), PARAM (blue) and ERA5 (red) at the selected timesteps from Figure \ref{fig:imerg_san}, namely at all 06 UTC and 18 UTC timesteps from 06 to 08 August 2012, expressed by the respective median (thick lines) and the extent of the minimum-maximum "ensemble" spread (shaded areas). The FSS level beyond which the product can be considered skillful ($FSS_{uniform}\sim0.53$) is marked by the horizontal black dashed line. It is evident that the median FSS curves of EXPLC cross $FSS_{uniform}$ at a smaller spatial scale than ERA5 and PARAM (crossing points at 2.25° vs. 3.25° and 4°, respectively). Up until $\sim$2.75°, EXPLC also shows a faster increase in FSS as the neighbourhood size increases compared to ERA5 and PARAM. These results indicate that the spatial error regarding the most intense vortex-related rainfall is smallest in EXPLC among all products. The slower increase of the median FSS curves for ERA5 and PARAM at scales below 2° suggest general shortcomings in the spatial placement of the heaviest vortex-related rainfall with parameterized convection. For PARAM, this is generally true across all timesteps considered (see blue shaded area). Eventually, the curves converge around 4.5°, beyond which little to no distinction can be made regarding the performance of the products. Furthermore, lowering the threshold to the 75th percentile does not close the performance gap between PARAM and EXPLC (not shown).

The performance of the products is further evaluated with SAL in Figure \ref{fig:fss_sal}b, where the S- and A-component are plotted on the abscissa and ordinate, respectively. The median values of both are denoted by the dashed lines as well as in Table \ref{tab:fss_sal} for all components. Eventually, the shaded areas show the degree of dispersion of the data points defined by the minimum and maximum values of the S- and A-component. The L-component of each timestep is depicted by the size of the data points. Recall that a value of zero indicates a perfect result for all components. In general, while the medians show a similar tendency for all products, distinctions between explicit and parameterized convection are notable. EXPLC tends to underestimate the sizes of the rainfall structures compared to IMERG (i.e., negative S-values), which is in line with presumptions made in section \ref{sec:sec4_san} that EXPLC generally struggles with convective organization. This does not seem to be prevalent with parameterized convection. Considering the A-component, EXPLC underestimates the rainfall amounts around the vortex, confirming the observations made in the previous section. However, the A-component variability in EXPLC is visibly smaller than for both PARAM and ERA5, partly since the latter show instances of strong overestimation of rainfall. Overall, while rainfall in EXPLC is generally too weak and small, both PARAM and ERA5 appear to have a higher ability to form organized rainfall. The L-component is not trivial to interpret, as it consists of two parts (see section \ref{sec:methods}). However, both PARAM and ERA5 are more prone to larger location errors (i.e., values around 0.5) than EXPLC, which is reflected in slightly higher median values (Table \ref{tab:fss_sal}). For ERA5, this is particularly interesting insofar as it shares the same center location of the 6°x6° domain as IMERG (see section 3.2). Therefore, the location errors with parameterized convection are likely the major source of lower skill according to the FSS as well.

\begin{table}[!htbp]
	\caption{The median values of S-, A- and L-values from comparing 3-hourly rainfall in EXPLC, PARAM and ERA5 with IMERG for the San and Kenieba cases. The S- and A-values are additionally represented in Figures \ref{fig:fss_sal}c and d,by the dashed lines.}
	\label{tab:fss_sal}
	\centering
	\begin{tabular}{lcccccc}
		      & \multicolumn{3}{c}{San case} & \multicolumn{3}{c}{Kenieba case}\\
              & S-Score & A-Score & L-Score & S-Score & A-Score & L-Score \\
	    EXPLC & -0.50 & -0.43 & 0.38 & -1.09 & -0.41 & 0.54 \\
	    PARAM & -0.32 & -0.03 & 0.44 & -0.37 & -0.14 & 0.22 \\
	    ERA5  & -0.14 & -0.21 & 0.48 & -0.16 & -0.20 & 0.32 
	\end{tabular}
\end{table}

\subsection{Kenieba case}

In same fashion as for the San case, Figures \ref{fig:fss_sal}c,d show FSS and SAL scores for the Kenieba case considering all 06 UTC and 18 UTC timesteps from 23 to 25 August 2019. As outlined in section \ref{sec:sec4_ken}, EXPLC struggles with the organization of convective clusters by producing rather scattered rainfall systems while PARAM captures the basic development well. Considering the median FSS, the rainfall fields in PARAM become skillful at a smaller scale than for either EXPLC or ERA5 (intersection of FSS curves with $FSS_{uniform}$ at 2.25° vs. 3° and 3.25°, respectively). However, considering the range of FSS curves, there is a large spread in the performance of EXPLC (see green shaded area). Since FSS is a spatial measure, the strong variation of FSS across all timesteps is most likely due to the spatially random nature of unorganized convection. In contrast, the FSS curves of PARAM exhibit much lower variance along the axis of the neighbourhood size, which is likely a result of the greater convective organisation of the precipitation field compared to EXPLC (cf., Figure \ref{fig:rr_hovmoller}f).

The SAL method in Figure \ref{fig:fss_sal}d further indicate the challenges of EXPLC to organize convection. Among all products, EXPLC shows the largest errors regarding the S-component (also see Table \ref{tab:fss_sal}). In general, the data points largely concentrate in the bottom-left quadrant, indicating that EXPLC systematically produces too small and too weak rainfall systems. The data points for both PARAM and ERA5 are located closer to the center, thus suggesting smaller amplitude and structural errors in the rainfall fields than EXPLC. Furthermore, between PARAM and EXPLC, PARAM evidently show considerably smaller L-values, which generally supports the results with FSS in Figure \ref{fig:fss_sal}c . 

Overall, the spatial verification demonstrates that the notion of using EXPLC over PARAM to improve the forecast of extreme precipitation in West Africa is strongly case dependent. As explored in section \ref{sec:weather_systems}, the moisture conditions are one major aspect where both cases differ considerably. The next section delves into the performance of ICON in reproducing the dynamical fields in order to explore the contrasting performances of EXPLC.

\subsection{Comparison of dynamical fields between ERA5 and ICON simulations}
\subsubsection{Wind field} \label{sec:sec6_windfield}

\begin{figure}[!htbp]
	\centering
  \includegraphics[width=12cm]{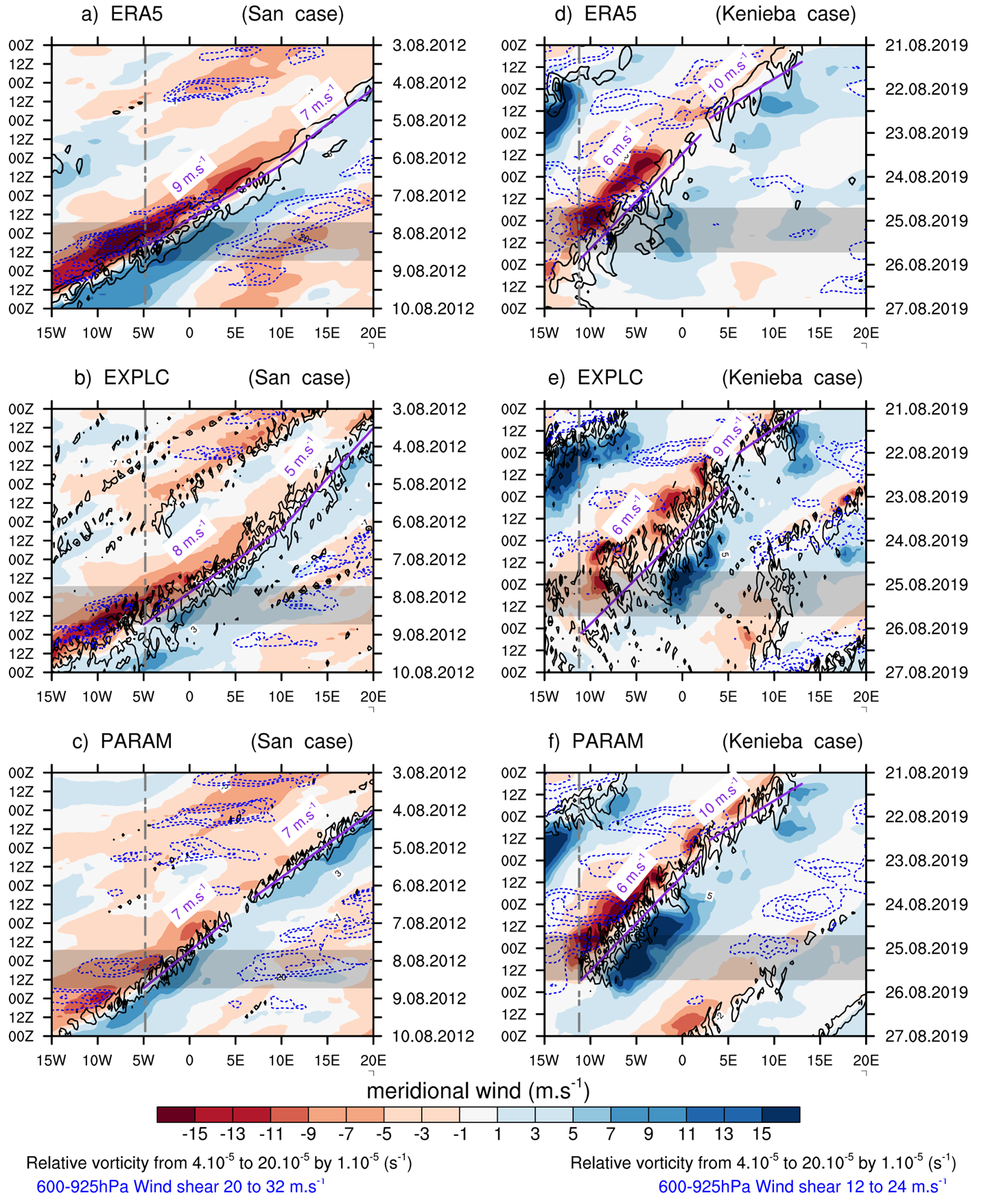}
	\caption{Hovmöller diagrams showing the evolution of 700 hPa relative vorticity (black contour), 700hPa meridional wind (shaded) and 925--600 hPa wind shear (blue contour) from ERA5 fields (a,d), EXPLC (b,e) and PARAM (c,f). The variables are averaged over a 5° latitude band centered around the coordinates of San and Kenieba, respectively. Violet lines indicate the position of maximum vorticity with the westward propagation velocity labelled. The panels on the left side depict the event of 08 August 2012 in San and panels on the right side the event of 25 August 2019 in Kenieba. The vertical lines indicate the longitudinal positions of San and Kenieba. The horizontal gray-shaded band refers to the time when the rainfall event was active according to IMERG.}
	\label{fig:wind_hovmoller}
\end{figure}

Both the San and Kenieba cases have shown the relevance of a cyclonic vortex in the respective development of extreme rainfall, the latter of which imposed challenges for ICON to capture it (see Figure \ref{fig:rr_hovmoller}). To investigate the dynamics associated with the cyclonic vortices, Hovmöller diagrams in Figure \ref{fig:wind_hovmoller} depicts the evolution of the meridional wind (colored shading), 700 hPa relative vorticity (black contours), and the 950--600hPa wind shear (blue dotted contours) based on ERA5 (Figure \ref{fig:wind_hovmoller}a, b), and the outputs of EXPLC (Figure \ref{fig:wind_hovmoller}c, d) and PARAM (Figure \ref{fig:wind_hovmoller}e, f) for both San (left column) and Kenieba (right column) cases. The variables are averaged over a 5° latitude band centered around the locations of San an Kenieba, respectively.

In general, a slanted dipole in meridional wind in the Hovmöller diagram, in addition to a streak of positive vorticity in its center, usually reflects the propagation of a vortex, which becomes apparent in both cases. For the San case, first signs of an established cyclonic circulation appear around 05 August at 15°E (Figure \ref{fig:wind_hovmoller}a), which coincides with the longitude of Lake Chad (cf., Figure \ref{fig:weather_san}a). From there, the vortex gradually intensifies, as indicated by the strengthening in the meridional wind field, and propagates westwards at a speed of around 9 ms$^{-1}$, which falls into the typical velocity range of AEWs \citep{reed1977structure,fink2003spatiotemporal}. Upon intensification, the vortex creates a zone of pronounced vertical wind shear (i.e., >16 ms$^{-1}$) at its western flank (i.e. within the region of northerlies) at around 07 August, which likely facilitated the intensification and organization of the squall line (cf., Figure \ref{fig:rr_hovmoller}a). Compared to ERA5, clear indications of a stable cyclonic circulation appear later in time in both EXPLC (Figure \ref{fig:wind_hovmoller}c) and PARAM (Figure \ref{fig:wind_hovmoller}e), exhibiting a slightly lower propagation velocity (7-8 ms$^{-1}$) than in ERA5. Moreover, both simulations are unable to produce a comparable area of strong vertical wind shear, which might partly explain the smaller degree of convective organization compared to IMERG. Nonetheless, the overall evolution of the strength of the vortex is reasonably captured in EXPLC, even though the structures in relative vorticity are noisier compared to both PARAM and ERA5. The noisier pattern is likely related to rainfall systems at smaller scales due to the explicit treatment of convection.

For the Kenieba case, the intensification of the cyclonic vortex, whose velocity is overall comparable to AEWs as well,  sets in roughly one day prior to the event in the evening of 23 August (Figure \ref{fig:wind_hovmoller}d), concurrent with the convective organizations of the rainfall systems (cf., Figure \ref{fig:rr_hovmoller}b). While PARAM appears to capture the overall structure of the meridional wind field, wind shear, and relative vorticity features reasonably well in space and time (Figure \ref{fig:wind_hovmoller}f), EXPLC struggles to realize a pronounced vortex towards the event (Figure \ref{fig:wind_hovmoller}e). Again, the noise in relative vorticity in EXPLC is likely linked to scattered convective cells, which never develop into an organized system over Kenieba (cf., Figure \ref{fig:rr_hovmoller}d). Eventually, the overproduction of small-scale rainfall features in EXPLC might have inhibited an efficient co-existence of organized convection and vortex.

\subsubsection{Moisture field}

\begin{figure}[!htbp]
	\centering
  \includegraphics[width=12cm]{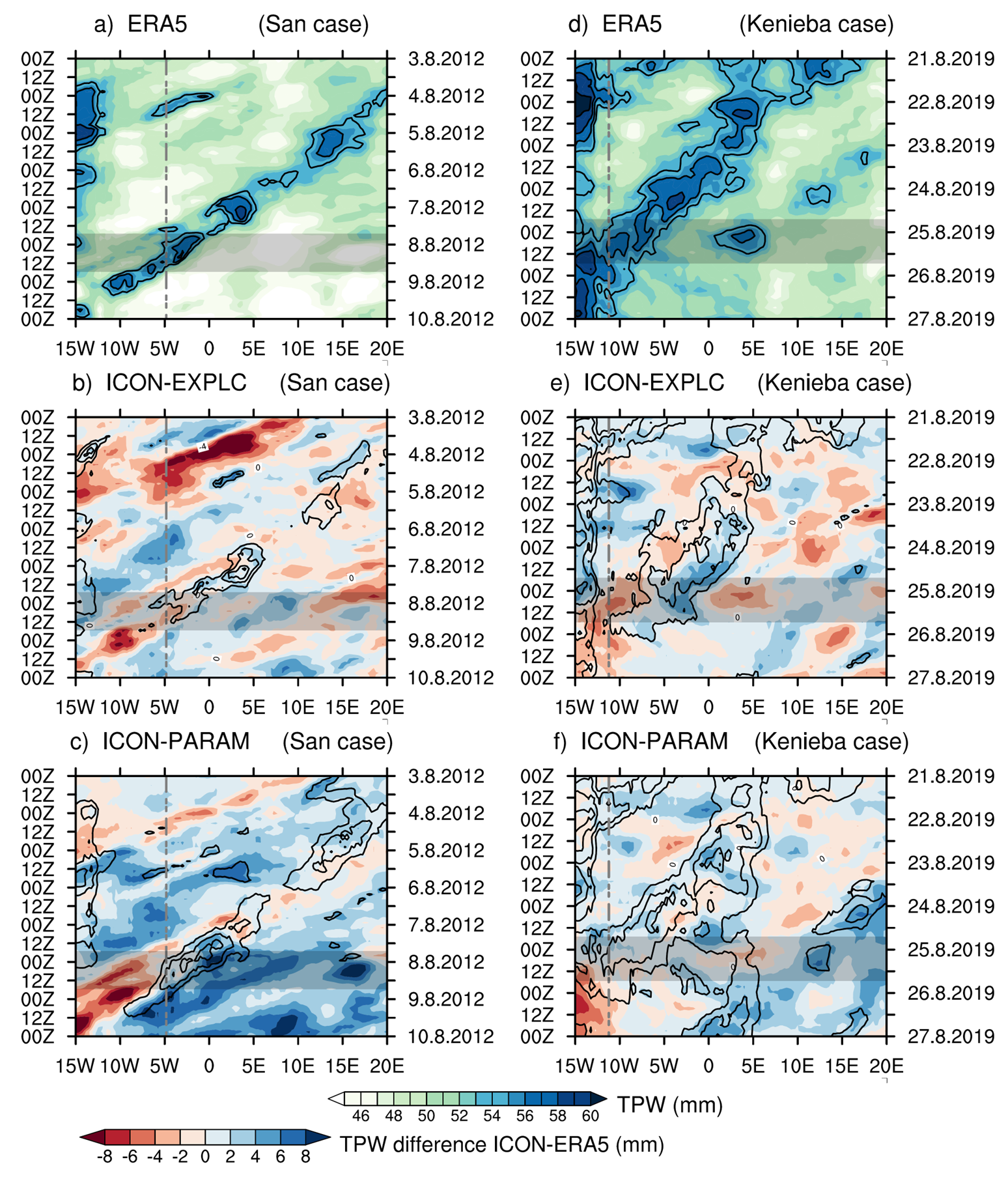}
	\caption{Hovmöller diagrams showing the propagation of total precipitable water (TPW) patterns from ERA5 (a,d), and differences in TPW between EXPLC-ERA5 (b,e) and PARAM-ERA5 (c,f). The variables are averaged over a 5° latitude band centered around the coordinates of San and Kenieba, respectively. The black contours show TPW values above 54 mm in 2 mm intervals. The panels on the left side depict the event of 08 August 2012 in San and panels on the right side the event of 25 August 2019 in Kenieba. The vertical lines indicate the longitudinal positions of San and Kenieba. The horizontal gray-shaded band refers to the time when the rainfall event was active according to IMERG.}
	\label{fig:pw_hovmoller}
\end{figure}

The Hovmöller diagrams in Figure \ref{fig:pw_hovmoller} depict the spatio-temporal evolution of total precipitable water (TPW) during the two extreme events in EXPLC and PARAM in comparison to ERA5. In both cases, enhanced tropospheric moisture content is clearly related to the respective westward moving cyclonic vortex (Figure \ref{fig:pw_hovmoller}a, b). It is also evident that the conditions across West Africa during the Kenieba case are overall moister than during the San case, the  reasons of which were outlined and analyzed in sections \ref{sec:sec6_weather_san} and \ref{sec:sec6_weather_kenieba}.

In the San case, EXPLC struggles to concentrate moisture within the vortex to the same level as ERA5 (Figure \ref{fig:pw_hovmoller}b), which might explain the underestimation in rainfall compared to IMERG (cf., Figure \ref{fig:rr_hovmoller}c). In this regard, PARAM performs better, but exhibits larger deviations in the overall moisture field (Figure \ref{fig:pw_hovmoller}c). Particularly the drying in the ridge before and after the AEW passage at San is strongly underestimated. The vortices in PARAM and ERA5 are slightly out of phase due to different propagation velocities (see section \ref{sec:sec6_windfield}), which largely explains the distinct westward propagating negative anomaly. In the Kenieba case, pronounced differences between EXPLC and PARAM are not as evident as in the San case. Again, PARAM tends to overestimate the moisture content in the area of and beyond the vortex (Figure \ref{fig:pw_hovmoller}f), whereas EXPLC is largely on the drier side (Figure \ref{fig:pw_hovmoller}d). 
Although it is difficult to draw general conclusions about model behavior from two cases alone, PARAM efficiently converts the abundance of moisture into strong precipitation within the vortex. For EXPLC, the Kenieba case hints at issues in moisture-laden conditions where the explicit treatment of convection in ICON produces too many small-scale rainfall systems too fast. This potential deficiency might be mitigated in situations like the San case which involves drier airmasses for convective enhancement.

\section{Summary and conclusion}  %% \introduction[modified heading if necessary]
\label{sec:summary}

Two extreme rainfall events in Mali during the peak monsoon season, namely the San case around 08 August 2012 and the Kenieba case around 25 August 2019, were investigated with respect to their rainfall structures and dynamical forcings, as well as the ability of DWD's ICON model to represent their evolution. With rainfall amounts of, respectively, 127 mm and 126 mm in 24 hours, these two cases feature different dynamic drivers that led to the intense rainfall. The daily precipitation accumulations are extreme (i.e., around or above the 99th percentile) in the respective station climatologies, the satellite-based IMERG dataset (final run, V6), and reanalysis data from ERA5. At a horizontal resolution of 6.5 km, ICON was used to perform two sets of simulations for each case: EXPLC and PARAM (ICON without and with convective parameterization, respectively) in order to test the capabilities of the two configurations to represent the extreme rainfall. Using a similar model set-up, these experiments were motivated by the study of \citet{pante2019resolving} who showed that EXPLC configurations generally exhibit a superior representation of Sahelian convection compared to PARAM. IMERG and ERA5 data were used to (a) analyze the extreme precipitation events from a dynamical point of view, and (b) evaluate the performances of the model simulations. The most important results are the following:

\begin{itemize}
    \item The two extreme events at San and Kenieba were caused by organized convection associated with a low-level cyclonic vortex, however each within different environmental settings: While both cases featured an African easterly wave (AEW), the San case involved convective enhancement along dry Saharan airmasses, whereas the Kenieba case occurred within an unusual widespread wet environment extending deep into the Sahel.
    %While the San case featured an AEW, the Kenieba case developed in an extremely widespread moist lower troposphere facilitated by vortex couples over the continent and the Atlantic.
    \item While EXPLC satisfactorily simulates the San case, PARAM is superior for the Kenieba case. EXPLC exhibits a better representation of AEW-related dynamics and rainfall in the San case -- albeit underestimating the overall rainfall amount -- but struggles with appropriate convective organization within the anomalously moist environment in the Kenieba case by producing too many scattered rainfall systems. Here, PARAM appears to be capable of converting the abundant moisture into excessive rainfall in a more efficient manner through the moist vortex. This indicates that the allegedly benefit of using explicit convection schemes does not always perform better in extreme rainfall cases in the Sahel -- at least with ICON.
    \item The spatial verification of the rainfall fields in ICON with the FSS confirmed the issues of EXPLC in the Kenieba case, which is skillful at larger spatial scales compared to PARAM around the time of the extreme events. Even the grid-scale precipitation fields in ERA5 exhibit more skill than EXPLC, suggesting that parameterized convection might cope better in moisture-laden situations with relatively weak dynamic triggers.
    \item The analysis with SAL confirms the negative bias in the size and amplitude (i.e., intensity) of precipitation objects in EXPLC for both cases. The fact that EXPLC struggles with convective organization and instead triggers randomly distributed smaller convective cells leads to large location errors, particularly in the Kenieba case. PARAM does not seem to suffer from such deficiencies as soon as the vortex creates large moisture content.
    \item The two cases were influenced by the presence of convectively favourable phases of both TD/AEW and MRG, i.e. fast equatorial wave modes. When overlapping, convective activity and intensity encountered a boost. Furthermore, the AEW during the San case belongs to the 10\% most intense in the 20-year August sample (2000--2019). ER as a slow wave mode likely had a low impact on the convective activity itself, but might have contributed to the moist conditions during the Kenieba case. These results are consistent with \citet{peyrille2023tropical}. 
\end{itemize}

While EXPLC led to an improved representation of the diurnal cycle of precipitation and the westward propagation of rainfall systems in the Sahel across a large sample of ICON simulations (e.g., \citet{pante2019resolving}), its performance compared to PARAM can still be highly case-dependent, as the present study suggests. It  highlights the potential challenges of EXPLC to form organized convection in situations with high low-tropospheric moisture content and weak dynamical triggers. Further test simulations with increased spatial resolutions, later initialization times closer to the extreme event, and a different dataset for initial and boundary conditions (i.e., ICON analyses) led to enhanced rainfall amounts, but did not substantially improve the lack of convective organization in ICON for the Kenieba case (not shown). 

In this regard, one aspect which has not directly been addressed in this study is the influence of soil moisture on the meso-scale dynamics of convection. \citet{klein2020dry} have shown that enhanced (zonally-oriented) soil moisture gradients as a consequence of a southward excursion of the ITD can generate convective instability by promoting moisture convergence and stronger meso-scale heat fluxes in the planetary boundary layer (PBL). While distinct soil moisture gradients prevailed in the San case, they were considerably less pronounced in the overall moist Kenieba case (not shown). The sensitivity of such different levels of soil moisture variability on the PBL and cloud development in ICON was extensively studied in \citet{han2019response} in largely idealized settings. In their large-eddy simulations, the authors found a tendency of more air parcels with high vertical velocities in the case of strongly heterogeneous (cf., San case) soil moisture fields compared to more homogeneous (cf., Kenieba case) fields, thus suggesting higher chances for intense deep convection to form. How this finding in ICON translates into "real-world" NWP scenarios like the present two case studies, especially with respect to convective organization, needs further in-depth exploration to potentially gain more insight into the behaviour of EXPLC.

Certainly, results for the San case in this study align with previous studies over West Africa \citep{peters2019different, bechtold2021convection}, showing good performance of the convection-permitting configuration to simulate the westward-propagating MCSs linked with a pronounced AEW. However, the Kenieba case is one example that stands in contrast to this consensus. Beyond the challenges highlighted for the Sahel, this result may also be of relevance for the densely populated Guinea Coast region, where moisture exists in abundance. Extreme rainfall events in this region involving a moist vortex and high flood risks are increasingly documented, as shown by the works of e.g., \citet{maranan2019interactions}, Vondou et al. (2025; submitted to Quart. J. Roy. Met. Soc.), and Toure et al. (2025; submitted to Mon. Weather Rev.), and thus stress the need for reliable forecasts.

Despite the limitations of the study (only one model used and for only two cases), the results open up study perspectives on the following questions: To what extent is EXPLC more suitable for AEW-related extreme precipitation? Is the ability of convective organization in EXPLC generally decreased in situations with widespread high tropospheric moisture load and weak dynamical triggers that lead to many isolated cells? The latter question can be critical insofar as comparable extreme events such as in Ouagadougou 2009 were likely preceded by an unusual Sahelian wet spell as well \citep{lafore2017multi,beucher2020simulation}. Overall, this study underlines the ongoing need for process-based evaluations of NWP models for the understudied West Africa to increase the body of knowledge about atmospheric drivers of intense rainfall, and to identify and improve on continuous shortcomings in the formulation of parameterizations and the overall behaviour of models with respect to (extreme) precipitation. Due to the complexity of the dynamics of the West African monsoon, the latter likely requires steps towards more stratified and targeted verification strategies, e.g. based on climate zones or dynamical features which carries predictive skill in forecasting rainfall, such as AEWs.
\section*{Acknowledgements}
This work was supported by the DAAD Climate Research Alumni and Postdocs in Africa - (climapAfrica) Programme. Acknowledgement to the International Science Programme (ISP/IPPS) through the "Laboratoire d’Optique de Spectroscopie et des Sciences de l’Atmosphère (LOSSA)" at the Faculty of Sciences and Technology for providing workspace and facilities to the first author during the postdoctoral period. Special thanks to the IMKTRO research group for hosting the first author during his stay at KIT. We thank Dr. Athul Rasheeda Satheesh of IMKTRO for providing the Figures \ref{fig:waves_hovmoller} and \ref{fig:amp_phase_total}. This research was partly supported by the Deutsche Forschungsgemeinschaft (grant no. SFB/TRR 165, “Waves to Weather”) and conducted within the subproject C2: “Statistical-dynamical forecasts of tropical rainfall”. Prof. Dr. Andreas H. Fink and Dr. Marlon Maranan acknowledge funding from the German Federal Ministry of Education and Research (BMBF) through the WASCAL Research Action Plan 2.0 (WRAP2.0) for FURIFLOOD project (Current and future risks of urban and rural flooding in West Africa – An integrated analysis and eco-system-based solutions; grant No. 01LG2086A) and also from the BMBF project NetCDA (German Academic Network for Capacity Development in Climate Change Adaptations in Africa; grant No. 01LG2301E).

\section*{Conflict of interest}
The authors declare that they have no conflict of interest.

\section*{Authors’ contributions}
S. Sanogo: Conceptualization, formal analysis, investigation, data curation, visualization, writing - original draft, writing - editing; M. Maranan: Conceptualization, formal analysis, investigation, writing - original draft, writing - editing; A. Fink: Conceptualization, review, writing - editing; B. Woodhams: Conceptualization, review, writing - editing; P. Knippertz: Review, writing - editing.

\section*{Data availability statement}
Stations rainfall data used in this manuscript were obtained from the Karlsruhe African Surface Station-Database (KASS-D) from the Institute of Meteorology and Climate Research of the Karlsruhe Institute of Technology, Germany and from the Mali national agency for Meteorology (Mali-METEO). The ERA5 data were obtained from the Copernicus Climate Change Service (C3S) Climate Data Store (https://cds.climate.copernicus.eu). IMERG data are available through the NASA earth data portal (https://disc.gsfc.nasa.gov/datasets/GPM\_3IMERGHH\_06/).

\printendnotes

\bibliography{Literature_Sanogo_etal}

\end{document}